\documentclass[10pt,leqno]{article}

\usepackage{amsmath,amssymb,amsfonts,amsthm,mathrsfs,verbatim} 
\usepackage{algorithm}
\usepackage[noend]{algpseudocode}
\usepackage{setspace}
\usepackage{authblk}
\usepackage{caption}
\usepackage{geometry}
\usepackage{graphics} 
\usepackage{graphicx}  
\usepackage{hyperref}
\usepackage{makeidx}
\usepackage{color}
\usepackage{bm}
\usepackage{natbib}

\newtheorem{theorem}{Theorem}[section]
\newtheorem{proposition}[theorem]{Proposition}

\theoremstyle{definition}
\newtheorem{remark}[theorem]{Remark}

\newtheorem{example}[theorem]{Example}
\newtheorem{assumption}[theorem]{Assumption}

\def\R{{\mathbb R}}
\def\N{{\mathbb N}}

\def\M{{\mathbb M}}
\def\x{\theta}   

\title{Consensus group decision making under model uncertainty with a view towards environmental policy making}

\author[a]{P. Koundouri}
\author[b]{G. I. Papayiannis}
\author[c]{E. V. Petracou}
\author[d]{A. N. Yannacopoulos}

\date{}

\affil[a]{\footnotesize School of Economics and ReSEES Laboratory, Athens University of Economics and Business, GR; Department of Technology, Management and Economics, Technical University of Denmark, DK; Sustainable Development Unit, ATHENA RC; Sustainable Development Solutions Network-Europe; Academia Europaea}
\affil[b]{\footnotesize Section of Mathematics, Department of Naval Sciences, Naval Academy, Piraeus, GR; Stochastic Modelling and Applications Laboratory, Athens University of Economics \& Business, Athens, GR}
\affil[c]{\footnotesize Department of Geography, University of the Aegean, Mytilene, GR}
\affil[d]{\footnotesize Department of Statistics and Stochastic Modelling and Applications Laboratory,  University of Economics \& Business, Athens, GR}

\begin{document}
\maketitle

\begin{quote}
\begin{center}
\emph{This paper is dedicated to Professor A. Xepapadeas on the occasion of his retirement from Athens University of Economics and Business, with friendship and admiration.}
\end{center}
\end{quote}

\begin{abstract}
In this paper we propose a consensus group decision making scheme under model uncertainty consisting of an iterative two-stage procedure and based on the concept of Fr\'echet barycenter. Each step consists of  two stages: the agents first update their position in the opinion metric space by a local barycenter characterized by the agents' immediate interactions and then a moderator makes a proposal in terms of a global barycenter, checking for consensus at each step. In cases of large heterogeneous groups the procedure can be complemented by an auxiliary initial homogenization step, consisting of a clustering procedure in opinion space, leading to large homogeneous groups for which the aforementioned procedure will be applied.
 The scheme is illustrated in examples motivated from environmental economics. 
\end{abstract}

\vspace{10mm}

\section{Introduction}

Group decision making is an important field  with interesting applications in various disciplines, among which environmental economics. Group decision,  often requires that all or the majority of agents in the group agree to a single proposal or opinion, i.e. consensus. This is particularly true in cases where there is no coercion involved in the implementation of  the decision made, so that the implementation of the decision depends on the good will, or rather the acceptance of the common decision by all members of the group. 

To make the discussion more concrete we consider the following generic situation: Assume that a group of agents, $G$, has to reach a common decision concerning policies regarding a future contingency $X$. Policies may refer for instance to the cost of abatement measures for protection against $X$, which clearly require the acceptance of a commonly acceptable estimate for the value of $X$ by every member of the group as well as the acceptance of a commonly acceptably discount factor. Typically, different member of the group will have different valuations for $X$, therefore report different costs for the adverse effects of $X$. Moreover, different members of the group will have different discount rates for calculating the present value of the future adverse effect $X$. As a result of the above, each member $i$ of the group $G$ will report a different value for a reasonable cost $C_i$ of abatement measures taken today so as to ease the future effect $X$. This means that unless the abatement cost $C$ proposed by the policy maker (upon which the proposed policy measures are priced) is carefully chosen so that it is finally acceptable by every member of the group (whose report of the cost $C_i$ deviates from $C$) it will not be acceptable by all group members, therefore the policy (unless coerced) will not be successful. 

The above example, introduces the important notion of consensus, an important concept in group decision making, which essentially means choosing a proposal for the common decision, on which every  member of the group (or its majority) will agree upon, even though their initial  positions (anchor positions) may deviate from that. Consensus decision making is very important in group decision making where coercion is not applicable, as an example one may consider climate change negotiations. An important role in group decision making is played by the mediator, an agent that introduces a proposal (based on some opinion different to,  but somehow combining those of each member of the group) places it to the attention of the group and hopes for consensus.

The aim of this paper is to add to the literature on consensus in group decision making by touching upon a theme that to the best of our knowledge has not been yet sufficiently addressed, which consists of the concept of the Fr\'echet barycenter as a consensus point and the effects of deep uncertainty and agent inhomogeneity on the group decision making process. We propose  a dynamic mechanism for consensus in a general metric space setting for the opinion space, based on the concept of the Fr\'echet barycenter as  a choice of proposals for the mediator and for opinion update for the group members, which takes into account, among others,  interactions between agents and their effect on opinion update and acceptance of proposals, inhomogeneity in the characteristics of the agents, and model uncertainty. A more detailed account of the state of the art on dynamical models for consensus in group decision making and the contribution and placement of the present work in this literature is provided in Section \ref{STATE-OF-THE-ART}. The proposed numerical scheme can act as a useful simulation tool for assessing the effects of factors such as inhomogeneity of the group composition or enhanced uncertainty towards future events, etc, on the success of the consensus process and its speed of convergence.

The structure of the paper is as follows: In Section \ref{STATE-OF-THE-ART} we present a short literature review of the field and place our contribution within the current state of the art in the field by highlighting its points of originality. In Section \ref{Barycenter-23-9} we introduce the notion of opinion space as a metric space and present solid argumentation towards the choice of the Fr\'echet barycenter as a possible proposal or opinion update. In Section  \ref{CEA} we introduce the proposed  dynamical consensus scheme and illustrate it with various examples, and in Section \ref{COMMON-SDR} we study an application in environmental economics, related to the determination of the social discount rate and the valuation of future projects under uncertainty.

\section{State of the art and aims and contribution of the present work}\label{STATE-OF-THE-ART}

\subsection{State of the art and brief literature review}\label{SOA-1}

Group decision making is a subject that has been extensively investigated (see for example the review paper \cite{perez2018dynamic} and references therein). Quoting this reference ``Consensus in group decision making requires discussion and deliberation between the group members with the aim to reach a decision that reflects the opinion of every group member in order to be acceptable by everyone. Traditionally consensus reaching is theoretically modelled as a multistage negotiation process which ends at agreement'' (\cite{perez2018dynamic}). This quotation indicates the following salient features in consensus decision  making:

(a) the 
need to define the concept of the mean in opinion space (this is what would be understood as the decision that reflects the opinion of every group member as quoted above), especially in cases where the opinion space is a complex space, e.g. a space of beliefs of models

(b) The need for a mechanism where group members update their opinions, either under the influence of other group members or by the pressure of time to reach a consensus and

(c) the proposal of a reasonable multistage procedure (i.e. an iterative process) at each stage of which the group members will update their positions, so that in due course consensus is reached.

Tasks (a) and (b) call for an appropriate way of aggregating opinions so that the concept of ``opinion that reflects the opinion of every group member'' can be defined in a reasonable way. Various suggestions have been made on that,  ranging from  e.g. \cite{degroot1974reaching} where a simple averaging of probability distributions with an appropriate choice of weights was introduced, to
quantile averaging techniques as proposed in \cite{lichtendahl2013better} or \cite{petracou2022decision} (where more general approaches were also introduced),  techniques inspired from Bayesian statistics (see e.g. \cite{basili2020aggregation} and references therein), methods based on the concept of aggregation operators and minimal cost consideration (see e.g. \cite{zhang2023consensus} and references therein) methodologies inspired by fuzzy measures theory (see e.g. the review \cite{herrera2014review} and references therein. An important strand of literature concerning opinion and preference aggregation, particularly popular in the economics community revolves around social choice theory related methodologies, emphasizing on an axiomatic framework, that the aggregation mechanism should abide to (see e.g. \cite{brown1975aggregation}, \cite{vincke1982aggregation}, \cite{skiadas1997conditioning}, \cite{gajdos2008representation}, \cite{may1954intransitivity} for a necessarily incomplete list and the references therein). Often, the axiomatic framework, while enticing can be unnecessarily rigid, leading to various non possibility results such as for example the celebrated Arrow impossibility theorem, hence making the consensus approach more difficult. While this approach is not pursued in this work, a connection of our contribution with it is mentioned in Section \ref{SOA-2} below.
A particularly popular choice for aggregation mechanisms is the so called utilitarian approach, in which some sort of weighted average of preferences or opinions is used  (see e.g. \cite{gollier2005aggregation,bhamra2014asset,jackson2014present,ebert2020weighted,heal2014agreeing} and references therein, for a very incomplete list of citations).  Various suggestions for opinion aggregation in the group of work mentioned in the beginning of this paragraph, while not directly related to this approach, are in fact compatible with the utilitarian approach, as they resort to some type of averaging for obtaining the aggregate opinion.

Task (c) has equally been extensively studied, starting from the contribution of \cite{degroot1974reaching} to more sophisticated dynamical system oriented approaches (see e.g.  \cite{amirkhani2022consensus}), dynamical mechanisms with varying degrees of realism and complication (see e.g.  \cite{perez2018dynamic}, \cite{gupta2017consensus} and references therein), approaches taking into account the effects of social interactions through e.g. social networks (see e.g. \cite{urena2019review}, \cite{li2022consensus} and references therein) etc. Clearly the above list of references is indicative and unavoidably incomplete.

Moreover, the problem of determining the consensus point is connected with numerous applications in various fields, where game-theoretic representations of the related optimal decision problems under study are possible or model uncertainty issue appears. Some indicative fields of application concern the study of taxation problems related to environment \cite{jorgensen2010dynamic}, design of pension schemes \cite{baltas2022optimal}, games related to pollution control \cite{kossioris2008feedback}, modelling and addressing hybrid or fuzzy systems and/or networks \cite{ozmen2017robust, weber2011modeling}, \cite{kropat2018fuzzy}, \cite{savku2022stochastic}, operational problems where robustification of the decision making process is required (see e.g. \cite{ozmen2023robust}, \cite{ozmen2011rcmars}) and wider logistic problems \cite{das2021multi, paul2022green}.

\subsection{Scope and contribution of the present work}\label{SOA-2}
Eventhough (dynamic) consensus group decision making has been an active field of research for at least the last 50 years, there is still an abundance of open problems as the increasing number of recent publications in the field indicates. Our work aims in contributing to this vast literature, on some issues stemming from the following observation: The widespread acceptance of a proposal upon which the final decision is made by all members of a group is made more difficult (if not impossible) by the following two factors:

(i) \emph{Group Heterogeneity}: If the group of agents that has to reach a common  decision has a widespread spectrum of positions in opinion space, i.e. presents large ``variance'' (with the concept of variance to be made concrete in Section \ref{24-9-OPINION} below), then the prospect of agreement to a common position is rather grim. A midpoint in position space has somehow to be proposed, so that bona fide agents willing to deviate from their initial positions in the interest of agreement, will not feel that their deviation is far larger than that of their counterparts.

(ii) \emph{Model Uncertainty}: If there is not a single model for $X$, to which all the agents in the group abide, then each agent may adopt a different model for $X$ and therefore report different estimates $C_i$ for the cost of $X$ (with a similar situation for the discount rates, see e.g. Section \ref{Gollier}). Hence, model uncertainty may contribute even more to group heterogeneity (see (i) above) and make group consensus even harder. These considerations introduce the need for choosing a commonly acceptable model for $X$, by the whole group, which will be subsequently used for valuation purposes, upon which policy making will be based. This is related again to the concept of the mean and variance in the space of models for $X$.

The aim of this paper is to address the question of group decision making with the above points (i)-(ii) and difficulties (which to the best of our knowledge have not yet been adequately addressed) in mind. In particular, we propose a scheme for consensus group decision making, in the presence of group heterogeneity and model uncertainty based on the modelling of the opinion space of the agents as an appropriate metric space, and the concept of the Fr\'echet mean  (barycenter) and variance. If connection with the economics literature on opinion and preferences aggregation is to be made, our approach hinges on the utilitarian approach (see Section  \ref{SOA-1}), suitably modified for the context under consideration. 
While we do not focus on an axiomatic framework, similarly to a large part of the literature on group decision making that does not consider the axiomatic framework at all, connection with an axiomatic framework is feasible, within the axiomatic framework  for variational utilities and model uncertainty, provided in the seminal work of \cite{maccheroni2006ambiguity}; this can be done through the concept of Fr\'echet variational utilities (see \cite{petracou2022decision}).
We report a two stage group decision making process that first identifies almost homogeneous groups of agents (in terms of opinions) hereafter called clusters, and then uses the representative opinions in the clusters for a proposal which is a candidate for common acceptance, in terms of the  barycenter of the representative opinions of each cluster. Moreover, we introduce the concept of learning i.e. we allow the agents to update their initial opinions (anchor points) as a result of interaction with their peers and propose an evolutionary process of opinion updating and proposal making (e.g. by the mediator) that results to consensus. While the proposed decision making process is of wider interest in group decision making, it is inspired and illustrated within the context of environmental economics, a field which accommodates all of the above mentioned features  (i) a feeling that we must agree, (ii) compliance to an agreement is voluntary and non coercive, hence relies on proposals that will be widely acceptable by all members of the group (iii) contingencies $X$ are subject to  model uncertainty and (iv) the decisions to be made are subject to great heterogeneity of the agents involved, due to their spatial scales. 

Among the contributions of the paper we report the following:

\begin{enumerate}
\item The conceptualization of opinion space as a very general metric space, allowing for the general treatment within a common framework of a large number of diverse situations (including situations where model uncertainty plays an important role).
\item The identification of the Fr\'echet barycenter as a possible consensus point for a group of agents, based on two different lines of argumentation: (a) A geometric approach  which identifies a Fr\'echet barycenter in the metric space of opinions as the common point in which all agents are comfortable with, with the least possible displacement in opinion space and (b) A probabilistic approach which identifies the Fr\'echet barycenter as the position in opinion space where the probability of acceptance by the group is maximized in a very general metric space framework\footnote{A  particular example of this result in the special case where the opinion space is the space of probability measures on $\R$ appeared in  previous work of some of the contributors of the present work \cite{petracou2022decision}; here this result is extended to the case of more general opinion spaces.}.
\item We propose a concrete opinion updating mechanism based on the concept of the Fr\'echet barycenter (see item 2 above), and the possible dependence structures between agents, with a view towards proposing a multistage negotiation process of opinion updating, that will eventually lead to consensus.
\item We propose and implement a dynamical algorithm for modelling the convergence towards consensus process, which allows us to assess both qualitatively and quantitatively the effects of (a) inhomogeneity of agents preferences (either in terms of variance in opinion space and/or differences in the time discounting  or the propensity of deviate from anchor positions) and (b) uncertainty or  (c) dependencies between agents on the possibility of consensus for the group or the time required to reach consensus. This dynamical algorithm can be turned into a powerful simulation tool for negotiation processes.
\end{enumerate}

The setting of the scheme is inspired by \cite{bishop2021network} where the consensus problem of a network of agents is considered and the convergence to a point is investigated for probability measures in the real line and existence results for weighting matrices that lead to a consensus are provided. However, the matter of how each agent chooses and reallocates at each step her/his weighting vector (resulting to the weighting matrix for all agents in the network) remains an open problem as stated in the same paper. In our work, we attempt to model the evolution of the adjacency matrix relying on standard behavioural aspects of the agents (e.g. desire a consensus to be achieved soon, desire to deviate from the anchor opinion, etc) combined with the notion of Fr\'echet barycenter. In this perspective, the discussed evolutionary scheme acts as a simulation and prediction mechanism for the consensus to be reached on a network of agents, subject to their preferences and behaviour. In fact, through numerical experiments it allows us to investigate how the different behavioural patterns and preferences heterogeneity affect the common consensus location and the time to agreement.

\section{Opinion space as a metric space and the Fr\'echet barycenter as consensus point}\label{Barycenter-23-9}

In this section we motivate the modelling of opinion space as a metric space $M$ endowed with an appropriate metric $d$, which serves as a measure of dissimilarity between the different opinions of the various agents involved. We also motivate  the concept of the Fr\'echet barycenter as a consensus point.

\subsection{Opinion space as a metric space}\label{24-9-OPINION}

Our fundamental  assumption is that  opinion space is modelled as a metric space i.e. a set endowed with an appropriate notion of distance or dissimilarity which will also allow for the quantification of variability between beliefs in the opinion space. In the abstract framework, to be made more concrete shortly, we assume that each agent $i$ carries an opinion (stand point) concerning the issue under consideration that can be considered as a point $x_i$  in some set $M$.  The dissimilarity between different opinions can be quantified in terms of a metric on $M$, i.e. a function $d : M \times M \to \R_{+}$ such that for any points $x_{i}, x_{j}, x_{k} \in M$ it holds that
\begin{itemize}
\item[(i)] $d(x_{i},x_{j}) \ge 0$ with  $d(x_{i},x_{j})=0$ if and only if $x_{i}=x_{j}$,
\item[(ii)] $d(x_{i},x_{j})=d(x_{j},x_{i})$,
\item[(iii)] $d(x_{i},x_{j}) \le d(x_{i},x_{k}) + d(x_{k},x_{j})$.
\end{itemize}
Adopting such a dissimilarity measure $d$ for any two opinions $x_i,x_j$ in $M$, the larger the $d(x_{i},x_{j})$ is, the greater the difference between these two opinions will be. 

The opinions of a group of $N$ agents as a collection of $N$ points $x_{i}$ of a metric space $(M,d)$, are collected in a set
$\M=\{ x_1, \ldots, x_N\} \subset M$. How can we define a notion of mean opinion for the group or any of its subgroups? Clearly such a notion would be very useful when trying to characterize the common trends in the opinion of heterogeneous groups of agents, or when we wish  quantify consensus. A common notion of mean, used in various applications in statistics or decision making (see e.g. \cite{degroot1974reaching}) is a linear estimator of the form $\hat{x}=\frac{1}{N} \sum_{i=1}^{N} x_{i}$, or more generally $\hat{x}=\sum_{i=1}^{N}w_{i} x_{i}$ for a choice of weights $w=(w_1, \ldots, w_{N})$. Such a choice may lead to an object $\hat{x}$, which cannot be identified as an element of the original opinion space $M$, in the case where $M$ is not a vector space, hence leading to a notion of mean that cannot be properly interpreted. In these cases, where $M$
does not carry a linear structure, an alternative definition for the mean must be  introduced. 

The appropriate choice in such cases is the notion of the Fr\'echet mean.  Given a choice of weights $w=(w_1, \ldots, w_{N}) \in \Delta^{N-1}$, (where by $\Delta^{N-1}$ we denote the $N$ dimensional simplex\footnote{$\Delta^{N-1}$ denotes the $N$-dimensional unit simplex, i.e. $\Delta^{N-1} := \{x\in\R^{N}\,\, : \,\, \sum_{i=1}^N x_i = 1, \,\,\, x_i \geq 0, \,\, \forall i\}$}, a measure of the variability of opinions in the set $\M$, can be given in terms of the function 
\begin{eqnarray*}
F_{\M} : M \to \R, \,\,\, F_{\M}(z):=\sum_{i=1}^{N} w_{i} d^{2}(z, x_{i}),
\end{eqnarray*}
which is called the Fr\'echet function of the set $\M$. The quantity
\begin{eqnarray}\label{VARIANCE}
V_{\M} := \min_{z \in M} F_{\M}(z),
\end{eqnarray}
is called the Fr\'echet variance of the set $\M$, and  its magnitude is a measure of the variability of elements contained in the set. The smaller $V_{\M}$ is the more homogeneous  the set is, while larger values of $V_{\M}$ indicate high heterogeneity in the set. Moreover, the minimizer of $F_{\M}$,
\begin{eqnarray}\label{MEAN}
z_{\M} := \arg\min_{z \in M} F_{\M}(z) = \arg\min_{z \in M} \sum_{i=1}^{N} w_{i} d^2(z,x_{i}),
\end{eqnarray}
is called the Fr\'echet mean \cite{frechet1948elements} or the Fr\'echet barycenter of $\M$. It is the analogue of the ``mean''  of $\M$, i.e. an element of $M$ (not necessarily an element of $\M$) that can be understood as the best approximation of the elements in $\M$. 

\begin{example}[Scenarios in environmental decision making]\label{WASSERSTEIN-METRIC-EX}
Environmental decision making is based on scenarios concerning the future development of quantities of interest, which are used for planning e.g. decision over abatement policies and measure, calculation of the cost of policies etc. Such scenarios, which are based on probabilistic models, are expressed as probability distributions for the quantity $Z$ of interest, which are often conflicting (a situation often referred to as deep  uncertainty). Different agents or experts may abide to different models, hence a natural candidate for opinion space is the space of probability measures. This is not a linear space, and can be metrized in various ways, one of the most popular being the Wasserstein metrizations, in terms of the class of Wasserstein metrics $W_{p}$ which for any choice of probability measures $x_1:=P_1$, $x_2:=P_2$ is  defined by
\begin{eqnarray*}
    W_p(P_1, P_2) = \bigg\{ \inf_{Z_1 \sim P_1, Z_2 \sim P_2} {\mathbb E}[| Z_1 -Z_2|^p] \bigg\}^{1/p},
\end{eqnarray*}
which clearly indicates that it is related to the error of prediction of a random variable $Z$ due to model misspecification (i.e. if $Z$ is modelled using $P_2$ whereas the true model is $P_1$). The choice of $p=2$, is a very common choice, becoming increasingly popular in statistics, machine learning and risk quantification (\cite{panaretos2019statistical}, \cite{papayiannis2021clustering}, \cite{papayiannis2018learning}, \cite{papayiannis2018convex}, \cite{petracou2022decision} ).
\end{example}

\begin{example}[Metric spaces of curves: Social discount term structure]\label{METRIC-CURVE}
An important example that spans a wide range of applications in environmental economics is in the valuation of future costs or income. Suppose that agents are to face a 
payoff (or loss) $X(t)$ at time $t$. The value $V(0)$ of $X(t)$ at time $0$ is given by $V(0)=X(t) e^{-r(t) t}$ where $r(t)$ is the discount rate between the time instances $0$ and $t$. The function $t \mapsto r(t)$ is important in the cost-benefit analysis of any project.  While integrable or continuous curves can be considered as elements of a vector space, the discount rate curves display specific characteristics, e.g. convexity or monotonicity which are important from the point of view of economics but at the same time make the set of social discount curves fail the properties of a vector space. 

As an example we propose the well known social discount term structure model of Gollier (see e.g. \cite{gollier2013pricing}), according to which the function $t \mapsto r(t)$ is of the form
\begin{equation}\label{sdr-model**}
\begin{aligned}
t \mapsto r(t)=r(t, y_{-1},\phi):=\delta+\gamma\frac{1}{t}\mu_{t}-\frac{1}{2}\gamma^{2}\frac{1}{t}%
\sigma_{t}^{2}, \\
\mbox{for} \,\,\,\,\,\,
\mu_{t}(y_{-1},\phi) :=\mu\,t+y_{-1}\frac{1-\phi^{t}}{1-\phi}, \,\,\,\,\, \\
\sigma_{t}^{2}(y_{-1},\phi):=\frac{\sigma_{y}^{2}}{(1-\phi)^{2}}\left[  t-2\phi
\frac{\phi^{t}-1}{\phi-1}+\phi^{2}\frac{\phi^{2t}-1}{\phi^{2}-1}\right]
+\sigma_{y}^{2}t,
\end{aligned}
\end{equation}
where $y_{-1}$ and $\phi \in [0,1]$ are appropriate parameters (details on how this model is obtained are provided in Section \ref{COMMON-SDR}). 
The set where possible discount rate curves live is then 
\begin{eqnarray}\label{R-set}
{\cal R}=\{ r \in C([0,T]) \,\, :  \,\, \exists \, (y_{-1},\phi) \in \R \times (0,1) \,\, \mbox{such that} \,\, \\
r(t)=r(t,y_{-1},\phi), \,\, \forall \,\, t \in [0,T] \}.  \nonumber
\end{eqnarray}
While ${\cal R}$ is a subset of the space of continuous functions (which is a vector space and a Banach space under a suitable norm) not any continuous function qualifies as a yield curve, from the point of view of economics, unless it has the particular shape  and qualitative properties described by the form of the functions in \eqref{R-set}. ${\cal R}$ is a nonlinear subset of $C([0,T])$, which can be described as a two dimensional manifold, it terms of its parametric representation in \eqref{R-set}.  Since linear combinations of curves in ${\cal R}$ will not result to a curve in ${\cal R}$, we need to consider ${\cal R}$ as a metric space, eventhough, it is embedded in the vector space $C([0,T])$. As an analogue to that, if you are an earthling your natural space is the surface of the globe (a sphere) and you are not allowed to consider in your motions the ambient vector space $\R^3$ (unless you risk finding yourself in the void).

\end{example}

\subsection{The Fr\'echet barycenter as consensus point}

The Fr\'echet barycenter of $\M$, for an appropriate choice of weights $w \in \Delta^{N-1}$ can  be a good candidate for the consensus point of the agents in the group.

\subsubsection{A geometric characterization of the barycenter as consensus point}\label{GEOM}

Consider $N$ agents and their opinions $\M=\{ x_{1}, \ldots , x_{N}\} \subset (M,d)$ and for simplicity assume that $M$ is compact (or a compact subset of a metric space). Each agent $i$ has a tendency to deviate around its central opinion (anchor), which can be modelled geometrically as follows: An opinion $x \in (M,d)$ will be considered as acceptable by agent $i$ as long as $d(x,x_i) \le \epsilon_i$. The larger $\epsilon_i$ is the more likely is agent $i$
to accept the opinion $x$ that does not coincide with her/his anchor point. The value of $\epsilon_i$ is a behavioural characteristic of the agent, modelling her/his firmness to initial opinion. A geometric interpretation of this is that an agent $i$ will accept an opinion $x\in M$ if it belongs to a ball centered at $x_i \in M$ of radius $\epsilon_i$, with the value of $\epsilon_i$ characterizing the agent.

We can now characterize a consensus point $x$ as the solution of the optimization problem
\begin{equation}\label{19-9-2023}
\begin{aligned}
\min_{x \in M, s \in R_{+}} s ,\\
\mbox{subject to}\\
d(x,x_i)^2 \le s, \,\,\, i=1, \ldots, N, \\
s \le \epsilon_{min}^2=\min\{ \epsilon_1^2, \ldots, \epsilon_N^2\}.
\end{aligned}
\end{equation}
The solution of problem \eqref{19-9-2023} will select this position in opinion space  that will require the least deviation from the anchor points of all agents, and is within the agreement ball of all agents.

\begin{proposition}\label{23-1}
The solution of problem \eqref{19-9-2023} corresponds to a  Fr\'echet barycenter of $\M$, for a selection of weights depending on $\epsilon_i$, and chosen so as to maximize a weighted version of the corresponding Fr\'echet variance.
\end{proposition}

The proof of Proposition \ref{23-1} is given in the appendix (see Section \ref{23-1-proof})

\begin{remark}
Problem formulation stated in \eqref{19-9-2023} considers as only plausible case the situation where all agents provide the same aversion preferences from their anchor points, i.e. $\epsilon_1 = \epsilon_2 = ... = \epsilon_N$. However, a more general representation of the problem could be provided by considering the case that each agent is allowed to deviate at different scale, i.e. considering different $s_i\geq 0$ for each agent $i$. In this case, problem \eqref{19-9-2023} could be replaced by
\begin{equation}\label{27-9-2023}
\begin{aligned}
\min_{x \in M, s \in R_{+}^N}, \sum_{i=1}^N \theta_i s_i, \\
\mbox{subject to}\\
d(x,x_i)^2 \le s_i, \,\,\, i=1, \ldots, N, \\
s_i \leq \epsilon_i^2, \,\,\, i=1, \ldots, N,
\end{aligned}
\end{equation}
for any weighting vector $\theta=(\theta_1,...,\theta_N)'\in\R^{N}$. This problem is treated similarly to \eqref{19-9-2023} by minor modifications of the relevant proof presented in the appendix \ref{23-1-proof}.
\end{remark}

\subsubsection{Barycenters as the most likely points of consensus}\label{26-9-0}

We now provide an alternative argument for the choice of the barycenter as the most likely agreement point of the agents in the group ${\cal G}$, with anchor points $\M=\{ x_{i}, \,\, i=1, \ldots, N\}$.
The basic assumption in this section is that the probability $p_i$ of an agent $i$ with opinion point $x_{i}$ to agree with an opinion $x$ depends on the distance $d(x,x_{i})$ of the proposal $x$ with the anchor point $x_i$. We assume that $p_i=\phi_{i}(d^2(x,x_{i}))$, where $s \mapsto \phi_i(s)$ is a decreasing function which models the agents propensity to deviate from her/his anchor position $x_i$ and accept a proposal $x$. The function $\phi_i$ models the agent's behavioural characteristics towards changes of position in opinion space.
As an indicative example of the choice of this function we offer the function 
$\phi_{i}(x) = \frac{1}{2} (1 + e^{ \alpha_{i} d^2(x,x_{i}) } )^{-1}$, $\alpha_i \ge 0$ which resembles the logistic model.

Assuming that the representative agents  are independent, we see that the probability of acceptance of proposed opinion $x$ by all the groups is equal to 
\begin{eqnarray*} 
P= p_{1} \cdots  p_{N} = \prod_{i=1}^{N} \phi_{i}( d^{2}(x,x_{i})).
\end{eqnarray*}
A reasonable choice for $x$, if common acceptance is required, is that $x$ which maximizes the probability of acceptance, i.e. the solution of the optimization problem
\begin{eqnarray}\label{max-agree-prob}
\max_{x \in M} P(x) = \max_{x \in M}  \prod_{i=1}^{N} \phi_{i}(d^{2}(x,x_{i})).
\end{eqnarray}

We will show that the solution of problem \eqref{max-agree-prob} corresponds to a Fr\'echet barycenter  for a choice of weights that depends on the functions $\phi_{i}$.

To simplify the proof, we make the following assumption.

\begin{assumption}\label{ASS-1}
Each element $x$ in the metric space $X$ can be parameterized in terms of a parameter $\theta \in H$, where $H$ is a suitable Hilbert space.
\end{assumption}

We emphasize the fact that the above assumption does not imply that $X$ is a vector space. The parameter space $H$ is a vector space but this does not mean the linearity of $X$. As an elementary example of that consider $X=S^{1}=\{ (x_1,x_2) \in \R^2 \,\, : \,\, x_1^2 + x_2^2 =1\}$, which is clearly not a linear space, but can be parameterized as $x_1=\sin(\theta), x_2=\cos(\theta)$, with the parameter $\theta$ living in the linear space $H=[0,2\pi]$. Other examples of more sophisticated opinion  metric spaces that satisfy Assumption  \ref{ASS-1} are the Wasserstein space of probability measures on $\R$, or location scale families of probability measures on $\R^n$ (see Example \ref{WASSERSTEIN-METRIC-EX}), as well as Example \ref{METRIC-CURVE}.

\begin{proposition}\label{23-2}
Assume that the functions $\phi_i$ are smooth. Then, every solution of problem \eqref{max-agree-prob} corresponds to a Fr\'echet barycenter of the set $\M$, with weights $w=(w_1,\ldots, w_N)$ determined by the functions $\phi_i$.
\end{proposition}

The proof of Proposition \ref{23-2} is given in the Appendix (Section \ref{23-2-proof}).

\begin{example}\label{NORMAL-BARY} Let ${\mathbb M}=\{P_1, \ldots, P_{K}\}$ with $P_{k} \in {\cal P}(\R^{d})$, and $P_{k} \sim N(\mu_{k}, S_{k})$, $\mu_{k} \in \R^{d}$, $S_{k} \in \R_{+}^{d \times d}$, $k=1, \ldots, k$. Then, a  solution of problem \eqref{max-agree-prob}  coincides with a barycenter of ${\mathbb M}$ with the weights $w=(w_1, \ldots, w_K)$ inherently determined by the anchor points and the preferences of the agents towards deviating from them. The details for the interested reader are provided in the appendix (see Section \ref{NORMAL-BARY-PROOF}).
\end{example}

\section{An evolutionary learning approach for reaching a consensus}\label{CEA}

In this section, an evolutionary framework, based on the concept of the Fr\'echet barycenter,  is proposed for the description of the behaviour for a number of agents when a consensus need to be reached, taking fully into account the heterogeneity of the agents and their dynamic interactions. 

\subsection{Motivation}

 The need for an evolutionary scheme for consensus achievement, should be obvious, however, as a motivation for our proposals we quote the following excerpt from \cite{perez2018dynamic}:
\begin{quote}
``\emph{Consensus in group decision making requires discussion and deliberation between the group members with the aim to reach a decision that reflects the opinions of every group member in order for it to be acceptable by everyone. Traditionally, the consensus reaching problem is theoretically modelled as a multi stage negotiation process, i.e. an iterative process with a number of negotiation rounds, which ends when the consensus level achieved reaches a minimum required threshold value. In real world decision situations, both the consensus process environment and specific parameters of the theoretical model can change during the negotiation period. Consequently, there is a need for developing dynamic consensus process models to represent effectively and realistically the dynamic nature of the group decision making problem.}''
\end{quote}

The above excerpt indicates the need for
\begin{itemize}
\item[(a)] Updating of opinions of the agents at each iteration based on influenced from their (possibly changing) environment
\item[(b)] The requirement of a moderator, that at each point of the procedure, will make a proposal, most likely to be acceptable by everyone, and ideally reflecting the opinions of all members, as much as possible.
\item[(c)] A criterion, or procedure for checking (e.g. by the moderator) whether consensus has been reached or not at the specific stage, or else the procedure will be repeated for a next round.
\end{itemize}
In all the above, the effects of agents inhomogeneity, uncertainty, time discounting effects and dependencies and interactions between agents must be accounted for.

The scheme that we propose takes into account all the points above making use of the analysis of Section \ref{Barycenter-23-9} to motivate an appropriate choice of an appropriate Fr\'echet barycenter, for the opinion update mechanism required  and the regulator proposal required in items (a) and  (b)  above, respectively. Metric spaces of probability models, will be well suited to  account for model uncertainty, and as stated in Section \ref{Barycenter-23-9} fall nicely within the Fr\'echet mean framework. The need for accounting for the interactions of the agents, which are important in opinion formation and may be dynamically changing will be accounted for by the adoption of a dynamic weighted graph, that models the agents dependencies, and the force of these interactions. The neighbourhood structure of the graph, i.e. the immediate dependencies of each agent, will play an important role in the weight selection procedure for the barycenter, chosen as the opinion update. As mentioned the proposal of the moderator at each stage must reflect the current opinions of the agents, reflecting the opinions of all agents (see item (b) above) and following the argumentation above, for this task, a Fr\'echet barycenter with equal weights is chosen. Finally for item (c), we choose to check agreement or not of each agent  to the moderator's proposal at each stage, by monitoring the distance of the proposal from the current position of the agent in opinion space, taking also into account the time preferences of the agents towards agreement.  This is consistent with our arguments in favour of the Fr\'echet barycenter as possible consensus (see Section \ref{26-9-0}).

\subsection{The evolutionary scheme}\label{23-9-100}

The proposed evolutionary scheme can be divided to the following stages: 

\subsubsection*{Initial Stage - Setting up a neighbourhood structure}

We first have to propose a network structure for the group of agents, which may model possible interactions, dependencies or affinities \footnote{which may affect the probability of acceptance of a proposal by an agent, depending on the acceptance or not of the proposal by other agents in the same clique} between them. To this end, we first consider a group of agents $G=\{1,2,...,N\}$ and a set of time-varying edges (links) $\mathcal{E}(t)$ formulating the time-varying graph $\Gamma(t)(G,\mathcal{E}(t))$.\footnote{Note here, that by agents we may either mean individual agents, or groups of agents, for example the clusters in opinion space $M$ obtained by the clustering procedure proposed in Section \ref{CLUSTER} (in which case each agent is identified with a cluster, so that $N=K$).} The set of neighbors of any agent $i=1,2,...,N$ is denoted by $\mathcal{N}_i(t) = \{ j\in G \,\,:\,\,(i,j)\in\mathcal{E}(t)\}$. 
The connectivity structure modelling the dependency of the agents is is expressed in terms of the  time-varying graph adjacency matrix $A(t)\in\R^{N\times N}$, whose elements $a_{ij}(t)$ are defined as 
\begin{eqnarray*}
\alpha_{ij}(t) = \left\{
\begin{array}{ll}
1, & i=j\\
1, & (i,j) \in \mathcal{E}(t)\\
0, & (i,j) \notin \mathcal{E}(t).\\
\end{array}\right.
\end{eqnarray*}
The graph can also be considered as a weighted graph, with time dependent weight matrix
$W(t) \in\R^{N \times N}$ representing the link intensity between the various agents. Standard assumptions that are made are: (i) $w_{ij}(t)\geq 0$ for any pair $(i,j)$ and any $t\in\mathbb{N}$ and (ii) $\sum_{j \in \mathcal{N}_i(t)} w_{ij}(t)=1$. Moreover, the aversion preferences the agents from their anchor opinions are determined by the parameters $\epsilon_i>0$ for $i\in G$, representing for each agent $i$ the radius of the maximum acceptable deviance from her/his anchor opinion.

\subsubsection*{Stage 1: Local updating of opinion for the agents} 
Assuming that at time $t$ the agents have not reached  a consensus and their opinions are identified by the set $\M(t) := \{ x_i(t), \,\, \forall i\in G \} \subset M$. At time  $t$, when the information concerning the current position of all agents is revealed, the agents re-allocate their beliefs in order to reach to a consensus in the  future. The time horizon in which each agent would like to reach consensus (so that the agreement is finalized) is subject to each agents time preferences and needs. Given that no consensus has been reached at time $t$, all agents enter a new round of negotiations, after renewing their original positions $x_{i}(t) \in M$ to a new position $x_{i}(t+1) \in M$ for all $i\in G$. In this position updating procedure, each agent $i$ is affected by her/his immediate neighbours ${\cal N}_{i}(t)$ in the time varying graph $\Gamma(t)$.\footnote{This is a reasonable assumption  since an agent's opinion is likely to be more affected by her/his immediate dependencies and/or pressure/interest groups.} Building on the results of Section \ref{Barycenter-23-9}  we propose that the new positions $x_{i}(t+1)$ for each agent $i$, is  a  local Fr\'echet barycenter of the points $\{x_{j}(t) \,\, : \,\, j \in {\cal N}_{i}(t) \}$ with an appropriate choice of weight  $w_{ij}(t)$ for each point in ${\cal N}_{i}(t)$, illustrating the interaction of the agent with her/his neighbours (influence, coercion etc). The selection of weights will be made by a weight update mechanism (see e.g. \eqref{weight-update}). 


In particular, the $i$-th agent's opinion is reallocated to the local barycenter
\begin{equation}\label{opinion-update}
    x_i(t+1) = \arg\min_{x \in M} \sum_{j \in \mathcal{N}_{i}(t)} w_{ij}(t+1) d^2(x, x_j(t)), \,\,\,\, \forall i\in G,
\end{equation}
while the weights are determined by the updating rule
\begin{equation}\label{weight-update}
\begin{aligned}
    w_{ij}(t+1) = \theta_i w_{ij}(t) + (1-\theta_i) R_{ij}(t), \,\,\,\, \forall (i,j) \in \mathcal{E}(t),
    \end{aligned}
\end{equation}
where 
$$ R_{ij}(t) :=  \frac{ \exp\left\{ - r_i d^2(x_i(t), x_j(t)) \right\} }{ \sum_{k\in \mathcal{N}_i(t)} \exp\left\{ - r_i d^2(x_i(t), x_k(t)) \right\} },$$
and $\theta_{i} \in [0,1]$ is an inertia parameter (representing the agent's tendency to persist in her/his current position) and $r_{i}$ models the agent's preferences towards reaching a consensus quickly.

\subsubsection*{Stage 2: Checking for consensus}  
Given their new positions $\M(t+1):=\{ x_{i}(t+1) \,\, : \,\, i \in G\}$ in opinion space, the agents check for consensus. This is done as follows: We form the global barycenter (e.g. with  homogeneous weights) 
 \begin{eqnarray*}
x_{B}(t+1) = \arg\min_{x \in M}\frac{1}{N} \sum_{i=1}^{N} d^2(x, x_{i}(t+1),
 \end{eqnarray*}
 and estimate, for each agent $i$, the probability of acceptance of the proposal $x_{B}(t+1)$. This probability of acceptance depends on the  distance of $x_{B}(t+1)$ from the current anchor point $x_{i}(t+1)$ and can be modelled e.g. as
\begin{equation}\label{prob-accept}
\begin{aligned}
q^{accept}_i(t+1) := P(\mbox{Agent i accepts proposal $x_B(t+1)$ at time t+1})\\ = e^{-\rho_i\,\, t\,\, d^2\left(x_i(t+1), x_{B}(t+1) \right)},
\end{aligned}
\end{equation}
where the sensitivity parameter $\rho_i>0$ models agent's $i$ propensity of deviating from her/his anchor position $\mu_{i}$. 

The probability $P$ of acceptance of proposal $x_{B}(t+1)$ by the group, is determined by the probability of acceptance of the proposal by the individual members $q_{i}^{accept}(t+1)$, either as a product if independence of agents is assumed or by implementing a dependence structure related to the graph $\Gamma(t+1)$. If $P$ is sufficiently high we stop else we return to step 1.
 

The evolutionary scheme is summarized in Algorithm \ref{alg-2}.
\begin{algorithm}[ht!]
\caption{The Evolutionary Consensus Learning Scheme}\label{alg-2}
\begin{tabular}{p{6cm}p{9cm}}
{\bf Step 0 (Initialization):}& Set $t=0$ and provide the initial beliefs $\mathbb{M}(0)$, the connectivity structure $W(0)$ and the preferences of each agent $i\in G$ through a parameter vector $\psi_i = (\theta_i, \rho_i, r_i, \epsilon_i)$.\\
{\bf Step 1 (Iteration Update):} & Set $t:=t+1$ and repeat Steps 2--5 till a consensus is reached.\\
{\bf Step 2 (Time-perspective update):} & Each agent updates her/his time preferences (if applicable, see e.g. criterion \eqref{time-update} in Appendix).\\
{\bf Step 3 (Connectivity Update):} & Each agent updates her/his local connectivity structure through criterion \eqref{weight-update}.\\
{\bf Step 4 (Opinion Update):} & Each agent updates her/his opinion (probability measure) through criterion \eqref{opinion-update}.\\
{\bf Step 5 (Acceptance condition):} & Each agent accepts the barycenter of the updated opinion set $\mathbb{M}(t)$ with probability of acceptance as determined in \eqref{prob-accept}. 
\end{tabular}
\end{algorithm}
The aspects that mostly affect the convergence to an agreement are expected to be: (a) the heterogeneity of beliefs and/or tendencies (propensities) of agents to update their anchor positions among the groups of agents, (b) the intensity of the connectivity and degree of dependence among the agents and (c) the  level of impatience of  each agent towards reaching   consensus, related to time discounting. These behavioural aspects are parameterized and introduced to the evolutionary procedure in the described scheme.

\subsection{A numerical experiment}\label{sec-3.2}

\begin{figure}[ht!]
\centering
\includegraphics[width=6in]{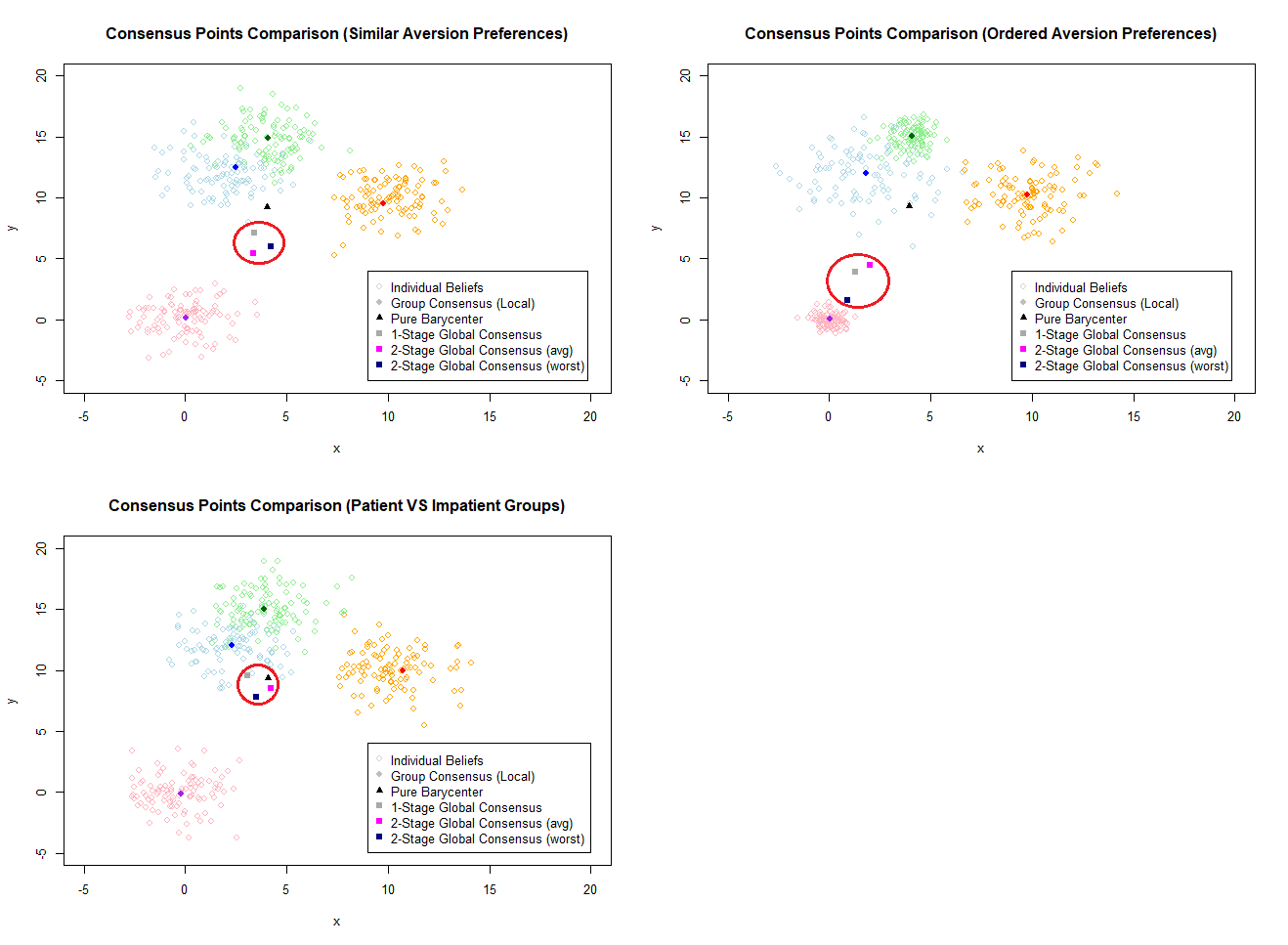}
\caption{Illustration of the agents' anchor opinions (different colour indicates different cluster), the local consensus points and the derived consensus points (marked in red) with the proposed evolutionary learning schemes for all three scenarios considered. }\label{fig-results}
\end{figure}

In this subsection we provide a numerical experiment employing the two consensus learning schemes described in the previous section to better understand and illustrate their behaviour and characteristics. Three different cases are considered concerning the agents' preferences and in particular are considered: (a) agents with similar aversion and time-discounting preferences, (b) agents with ordered preferences and (c) agents with different types of time-discounting preferences. To compare the required time for the one-stage and two-stage procedures we consider four different groups of agents where within each group there exist a homogeneity concerning the agents' preferences while between the groups the homogeneity level depends on the scenario. The one-stage scheme will handle all groups as one, while the two-step approach will first recover the groupings and then will apply the evolutionary method first within the groups and then globally to determine the consensus point. We consider elliptical groups with respect to the anchor opinions while the agents preferences within each group are generated in terms of the parameter vectors $\psi_{k,i} = (\theta_{k,i}, \rho_{k,i}, r_{k,i}, \epsilon_{k,i})$ where
\begin{eqnarray*}
 \theta_{k,i} \sim U([\theta_{L,k},\theta_{U_k}]), \,\,\,\, \rho_{k,i} \sim U([\rho_{L,k}, \rho_{U,k}]),\\
 r_{k,i} \sim U([r_{L,k}, r_{U,k}]),\,\,\,\, \epsilon_{k,i} \sim U([\epsilon_{L,k}, \epsilon_{U,k}]),
\end{eqnarray*}
for any $i\in G_k$ for $k=1,2,3,4$. The lower bound values $\theta_{L,k}, \rho_{L,k}, r_{L,k}, \epsilon_{L,k}$ and the upper bound ones $\theta_{U,k}, \rho_{U,k}, r_{U,k}, \epsilon_{U,k}$ differ per group $k$ depending on the scenario that is chosen. In Table \ref{tab-scenarios} are briefly summarized the scenarios to be considered in the simulation experiments and the preferences specification for each group. An illustration of the initial anchor preferences of all agents and the obtained consensus points by the one-stage and two-stage schemes are presented in Figure \ref{fig-results} while the required time steps till the derivation of the consensus points by all methods are displayed in Table \ref{tab-times}.

\begin{table}[ht!]\small
	\centering
	\begin{tabular}{l|l|cccc}
		\hline\hline
		{\bf Scenario} &  {\bf Agents' Preferences} & {\bf Group A} & {\bf Group B} & {\bf Group C} & {\bf Group D}\\ 
		\hline
		Similar preferences & Anchor opinion aversion & medium   & medium     & medium     & medium \\
		                                  & Time-discounting type   & indifferent & indifferent & indifferent & indifferent\\
		                                  &  Weighting Inertia effect & medium   & medium     & medium     &  medium\\
		 \hline                                 
		
		Ordered preferences & Anchor opinion aversion  & low        & medium & medium    & high\\
											 & Time-discounting type    & patient  & patient   & impatient & impatient\\
											 &  Weighting Inertia effect  & high      & medium & medium    & low\\
		\hline
	
		Patience VS Impatience & Anchor opinion aversion & medium& medium& medium& medium\\
										         & Time-discounting type    & patient & patient& impatient & impatient\\
										         &  Weighting Inertia effect  &     high &      high&   medium &           low\\
		\hline\hline	
	\end{tabular}
\caption{Description of each scenario considered for all agents and for each group}\label{tab-scenarios}
\end{table}

\begin{table}[ht!]\small
	\centering
	\begin{tabular}{lrrr}
		\hline\hline
		Scenario & One-Stage Scheme & Two-Stage Scheme (avg) & Two-Stage Scheme (worst)\\
		\hline
		Similar Preferences        &  89 & 85 (58) & 81 (54) \\
		Ordered Preferences     & 127 & 34 (27) & 57 (50) \\
		Patience VS Impatience &  79 & 45 (19) & 65 (39) \\
		\hline\hline
	\end{tabular}	
\caption{Time steps required for each scheme to derive the consensus point. In parentheses are displayed for the two-step schemes the time steps required to reach the local (cluster) consensus points.}\label{tab-times}
\end{table}

The employed methods seems to provide quite close consensus points in all scenarios considered. It is also evident that the two-step procedures are quite faster and since a part of the total steps are performed only with the $K$-fictitious agents, the complexity is quite lower than the appeared one. The pure barycenter is depicted in all three scenarios to realize the effect of the agents' preferences in the final agreement point. This is quite obvious in the second scenario (ordered preferences) where the pure barycenter is quite distant from the calculated consensus points by the methods. 

The numerical results of the algorithm indicate that large inhomogeneity of the agents may result to delays on the convergence of the algorithm. A way to bypass such problems would be to enhance the proposed method by including a preliminary stage in which a clustering procedure in opinion space is performed, leading to more homogeneous groups, on which we may apply the evolutionary procedure described above. This and other extensions are presented in  the Appendix, see Section \ref{EXTENSIONS}.

\section{Application in Environmental Economics: Convergence to a Common Social Discount Rate}\label{COMMON-SDR}

\subsection{Motivation}\label{motivation}

Climate change seems to be a common threat and consequently a dominant scientific and political concern and in high priority in the global agenda. It constitutes one of the most crucial problems that needs urgent cooperative negotiations and solutions in order to achieve agreements dealing with various bad consequences of our ways of life as well as production and consumption! The United Nations Framework Convention of Climate Change, the Kyoto Protocol and Paris Agreement are indicators for international political actions and negotiations to deal with impact of climate change.
Scientific knowledge for causes and effects of climate change and climate change’s economic and social impact worldwide are closely connected in the terms of Intergovernmental Panel on Climate Change with the goal to assess the global situation and recommend potential adoption of policies. 
Climate change is a multifaceted and complicated (it is not the only!) phenomenon which among others is related to international relations,  global governance in a geographically different and unequal world. Furthermore, it affects individuals and collectivities with uneven ways and with different levels of responsibility. In additions to power relations, climate change itself but also introduction and implementation of policies are related with present and future situations. Consequently, there is a need for common action!  Actually, causes, conditions, impact are different spatially and timely but they are assembled under the processing of capitalist organizing of way of life.
On various issues such as responsibility, justice, recommendation of policies and from whom and what are few issues of  debates. One of the important aspect of these climate change negotiations is whether we have achieved  consensus  and for what- scientifically – politically and on what.
Consensus is a wide issue/ element of negotiations and refers to different levels  such as social, political, economical, technical etc. as well as the time of intervention such as how urgent must be the actions, when, where, which are the institutional arrangements and in which direction – market, technology…
However, we have to consider about climate change’s causes  and crisis in order to identify potential conflicts and ways that we can overcome them.
Besides debates and disagreements scientifically, geographically and politically, consensus is important but also the involvement of a  mediator is a convenient way to overcome disagreement, scientifically and most importantly politically! 
If we would like to define processes of decision making, we must take into consideration procedural injusticies in the climate negotiations. 

The important factors in evaluating future contingencies  are (a) the social discount yield curve (SDR), providing the discount rate $r(t)$ by which a contingency $X(t)$ that is to be encountered at time $t$ needs to be priced at time $0$, and (b) an estimate of the probability distribution of the contingency $X(t)$, that will allow for the estimation of the contingency's value. Given these two, one can perform a valuation of the contingency as $K(0,t)={\mathbb E}[e^{r(t) t} X(t)]$. As typically the discount factor $r(t)$ depends on the time horizon $t$ at which the contingency will be encountered, a useful tool in valuation is the function $t \mapsto R(r)$ called the yield curve. Then, differing opinions can be modeled as elements of the vector space $M_1 \times M_2$, where $M_1$ is a space of yield curves (representing different views on the discount factors) and $M_2$ is a space of probability measures (representing different views on the distribution of future contingencies), metrized accordingly. These spaces were briefly introduced in Examples  \ref{METRIC-CURVE} and \ref{WASSERSTEIN-METRIC-EX} respectively, while more technical details are provided in Appendix E.

\subsection{Gollier's model for social discounting}\label{Gollier}

The social discount rate (SDR) is one of most fundamental parameters in cost-benefit analysis and its determination is of crucial importance in any valuation study or for policy making (see e.g. \cite{stern2007economics, nordhaus2007review},  \cite{gollier2002discounting, weitzman2007review, dasgupta2008discounting, heal2009climate, groom2005declining,hepburn2007recent,groom2007discounting,groom2007discounting1,hepburn2009social,koundouri2009introduction} for areas related to climate change) or environmental economics. The results of any valuation study are very sensitive to the choice of the social discount rate, and this sensitivity becomes more pronounced when longer horizon projects (such as for example environmental projects) are considered. Moreover, there is not unanimous agreement concerning the choice of the SDR, even when its calculation is based on widely accepted models, such as for example the Ramsey discounting formula.  

As an example of how controversies concerning the determination of  the SDR may arise between different agents, even when a single model is used, and its effects on the term structure of the discount rate we present the well known model for the determination of the SDR by Golier \cite{gollier2013pricing}, based on the classical Ramsey discounting formula. This formula connects the SDR with expected utility of consumption in the future in terms of
\[
r(t)=\delta-\frac{1}{t}\ln\frac{{\mathbb{E}}[u^{\prime}(C(t))]}{u^{\prime
}(C(0))}.
\]
In the above, 
\begin{itemize}
\item $r(t)$ is the discount rate at time $0$ for any contingency $X$ to be faced at time $t$
\item $\delta$ is the utility discount rate, 
\item $C(t)$ denotes consumption at time $t$ (a random variable unknown at time $0$) and 
\item $C(0)$ denotes today's consumption. 
\end{itemize}
From this formula, a term structure for $r$ is derived (i.e. the dependence $t \mapsto r(t)$), and is a crucial parameter in standard cost-benefit analysis (\cite{gollier2013pricing}).
For example, given the term structure, the cost at time $0$ of any contingency $X(t)$ to be faced at time $t$, is to be evaluated at $K(0,t)={\mathbb E}[e^{-r(t) t} X(t)]$, a formula which clearly indicates the sensitivity of the estimated cost, and hence any valuation or cost-benefit analysis based policy,  on the discount rate.

However, the future consumption  at time $t$, $C(t)$, is unknown at time  $0$ that $r(t)$ is to be determined. Hence, the determination of $r(t)$ requires estimates of future consumption, a quantity which may well be subject to the effects of model uncertainty. Consequently, this uncertainty is moved on to the discount rate term structure, and from that to any valuation. As a result of such uncertainty it is conceivable that for a group of agents, possibly having different beliefs concerning $C(t)$, there will be different opinions regarding $r(t)$ and for any valuation for contingencies $X$.

To make the arguments more concrete, let us follow Gollier's model (\cite{gollier2013pricing}) for the determination of the terms structure $t \mapsto r(t)$. We assume that a standard CRRA utility function with relative risk aversion $\gamma$ is used to value consumption. Moreover, the consumption process $C(t)$ follows a single factor (autoregressive) model of the form
\begin{equation}
\begin{aligned} C(t+1) & =C(t)\exp(x(t)),\\ x(t+1) & =\mu+y(t)+\varepsilon_{x}(t),\\ y(t) & =\phi y(t-1)+\varepsilon_{y}(t), \end{aligned}\label{25-3-2016-MOD}%
\end{equation}
where $\varepsilon_{x}(t),\varepsilon_{y}(t)$ are independent and serially independent with ${\mathbb{E}}[\varepsilon_{x}(t)]={\mathbb{E}}[\varepsilon_{y}(t)]=0$ and $Var(\varepsilon_{x}(t))=\sigma_{x}^{2}$, $Var(\varepsilon_{y}(t))=\sigma_{y}^{2}$, $y_{-1}$ is some initial state, and $\phi\in\lbrack0,1]$ is a parameter representing the degree of persistency (mean reversion) of $y$.  This model is supported by empirical data (see e.g. \cite{bansal2004risks}). Depending  on the value of $\phi$ the model can be either reduced to a standard random walk model which is a discretization of a Wiener process ($\phi=0$) or correspond to a discretization of an Ornstein-Uhlenbeck process ($\phi \ne 0$). Typically, $\{y(t)\}$ is an unobserved stochastic factor, which affects the observed growth rate $\{x(t)\}$ of the consumption process $\{C(t)\}$. Given values for $\phi$ and $y_{-1}$, the
stochastic consumption process $\{C(t)\}$ is lognormally distributed and in particular
\[
\ln C(t)-\ln C(0)\sim N(\mu_{t},\sigma_{t}^{2}),
\]
where
\begin{equation}\label{YYY}
\begin{aligned}
\mu_{t} =\mu\,t+y_{-1}\frac{1-\phi^{t}}{1-\phi}, \,\,\,\,\, \\
\sigma_{t}^{2} =\frac{\sigma_{y}^{2}}{(1-\phi)^{2}}\left[  t-2\phi
\frac{\phi^{t}-1}{\phi-1}+\phi^{2}\frac{\phi^{2t}-1}{\phi^{2}-1}\right]
+\sigma_{y}^{2}t.
\end{aligned}
\end{equation}
Using the general class of CRRA utilities, Gollier produces an analytic
formula for the term structure of the discount rate as
\begin{equation}\label{sdr-model}
r(t)=\delta+\gamma\frac{1}{t}\mu_{t}-\frac{1}{2}\gamma^{2}\frac{1}{t}%
\sigma_{t}^{2}.
\end{equation}
Note that in the above formula, the term structure is increasing or decreasing depending on the sign of $y_{-1}$. Moreover, in the case where $\phi=0$, the term structure is flat whereas for certain values it may have a convex structure. When all the parameters involved in model\eqref{25-3-2016-MOD} are fully known  the Ramsey formula can be used to produce a term structure for the SDR. However, even in this case the quantitative and qualitative (e.g. shape) properties of the term structure depend on the values of the parameters of the model, which themselves are not uniquely determined in terms of the available data. A calibration was performed in \cite{bansal2004risks} for  the factor model \eqref{25-3-2016-MOD} for consumption using annual USA  data  from the period 1929-1998, yielding the estimated parameters (monthly estimates) 
$$\mu=0.0015, \,\,\, \sigma_{x}=0.0078, \,\,\, \sigma
_{y}=0.00034.$$ 
 On the same work, the mean-reversion parameter was estimated to $\phi = 0.979$. Of course, these estimations are subject to statistical errors which allow for other valued of these parameters, compatible with the available data, that may lead to different models for $C(t)$ and subsequently different (both in a quantitative and qualitative sense) models for the term structure of the discount rate as provided by   \eqref{sdr-model}.

\subsection{Consensus achievement on the SDR and the probability model concerning the contingency: A numerical study}

Motivated by the discussion in the previous section, we devise the following gedanken experiment concerning consensus achievement on the SDR (and hence on the valuation of any contingency) by a group of agents who albeit all abiding to model \eqref{sdr-model} (with $C(t)$ provided by \eqref{25-3-2016-MOD}). The agents may have as anchor points versions of the model with different parameter values, hence resulting to different term structures for the discount factor and as a result different valuations of the same contingency $X$. The difference in the parameter values adopted by different agents in the group may arise from various reasons, among which being choice of different parameter values within the confidence interval for the US data, or the fact that different agents reflect different spatial locations and interests, i.e. are forming their time preferences for $r(t)$ in terms of future consumption for economies different than the US (hence leading to alternative calibrations for model \eqref{25-3-2016-MOD}).

For the simulation study, a group $G$ of $N$ agents is considered, with each agent reporting a different term structure curve for the discount rate $t \mapsto r_{i}(t)$, all collected in a set ${\mathbb M}=\{r_1(\cdot), \ldots, r_N(\cdot)\}$ of term structure curves. The set of curves ${\mathbb M}$ can be considered as a subset of a suitable metric space $M$, which will be chosen as a space of curves on $[0,T]$. This metric space of curves will serve as the opinion space $M$ in the context of Section \ref{24-9-OPINION} (and in particular Example \ref{METRIC-CURVE}). Technical details on the choice of metric and the calculation of the Fr\'echet mean in this context are provided in Appendix E. The agents in $G$ need to reach to a consensus towards the adoption of a commonly acceptable discount rate curve $r(\cdot)$ that will serve as the common instrument for the valuation of future contingencies $X$. Moreover, the group $G$ consists of three different subgroups (i.e. $G = G_1 \cup G_2 \cup G_3$) with each subgroup introducing different homogeneity levels concerning the agents' preferences. For instance, each subgroup could be realised as a different region of the world where different range of elasticities related to consumption are observed due to cultural divergences. The evolutionary algorithm introduced in Section \ref{CEA} is employed for exploring potential consensus points in the metric space $M$ of term structure curves, and investigate the dynamics of reaching consensus under the various heterogeneity levels between the agents with respect to the discount rate curves. The consensus discount rate curve, once and if reached, will be chosen as the SDR curve, used for the group $G$ and will be the outcome of the agreement, henceforth chosen for evaluating a certain contingency $X$.

The generation of the SDR curves $r_i(\cdot)$, $i=1, \ldots, N$, that the set ${\mathbb M}$ consists of, is made in accordance to the model of Gollier, presented in Section \ref{Gollier} by assuming that all agents abide to model  \eqref{sdr-model} (with $C(t)$ provided by \eqref{25-3-2016-MOD}), but adopting different values for the relevant parameters. To generate the opinion set ${\mathbb M}$ we sample a distribution of parameters for model \eqref{25-3-2016-MOD}, and then use \eqref{sdr-model} to generate the relevant discount rate curves. In particular, three different ranges for the elasticity parameter $\gamma$ are considered, and specifically $\gamma_1 \sim \mathcal{U}([0.8, 1.5])$, $\gamma_2 \sim \mathcal{U}([0.4, 1.7])$ and $\gamma_3 \sim \mathcal{U}([0.3, 2.0])$ (where subscript denotes the subgroup) representing different behaviours and heterogeneity on the agents' perspectives, while the parameters $\delta, \phi, y_{-1}$ are kept close to the calibration performed in \cite{bansal2004risks}, to capture more general behaviours. Specifically, these parameters are Uniformly and independently sampled as
$$ \delta \sim \mathcal{U}([0.029, 0.031]), \,\,\,\,
\phi \sim \mathcal{U}([ 0.977, 0.981 ]), \,\,\,\,
y_{-1} \sim \mathcal{U}([ -0.001, 0.001 ]).$$
According to this simulation scheme, each agent in $G$ will report a discount rate curve corresponding to \eqref{sdr-model}, with $C(t)$ generated by \eqref{25-3-2016-MOD} (equiv. \eqref{YYY}) with parameters $\gamma$, $\delta$, $\phi$ and $y_{-1}$ chosen as a sample point from the above distribution. This concludes the construction of the opinion set ${\mathbb M}$. The parameters related to the consensus process, i.e. the parameters determining the agents' determination and impatience to reach a consensus, are generated according to the simulation scheme described in Section \ref{sec-3.2}. For the simulation task a total number of $N=90$ agents is generated, with each subgroup consisting of $30$ agents. For the consensus determination task, both the one-stage and the two-stage processes are employed to illustrate and discuss the potential differences between the achieved consensus points.  Moreover, two different scenarios are considered concerning the agents' preferences: (a) the {\it Uniform Beliefs} scenario, under which the agents in all groups are assumed to display uniformly distributed preferences in reaching a consensus, and (b) the {\it Impatient Agents} scenario, under which agents of different subgroup display different patience levels on reaching a consensus.  

\begin{figure}[ht!]
	\centering
	\includegraphics[width=4.7in]{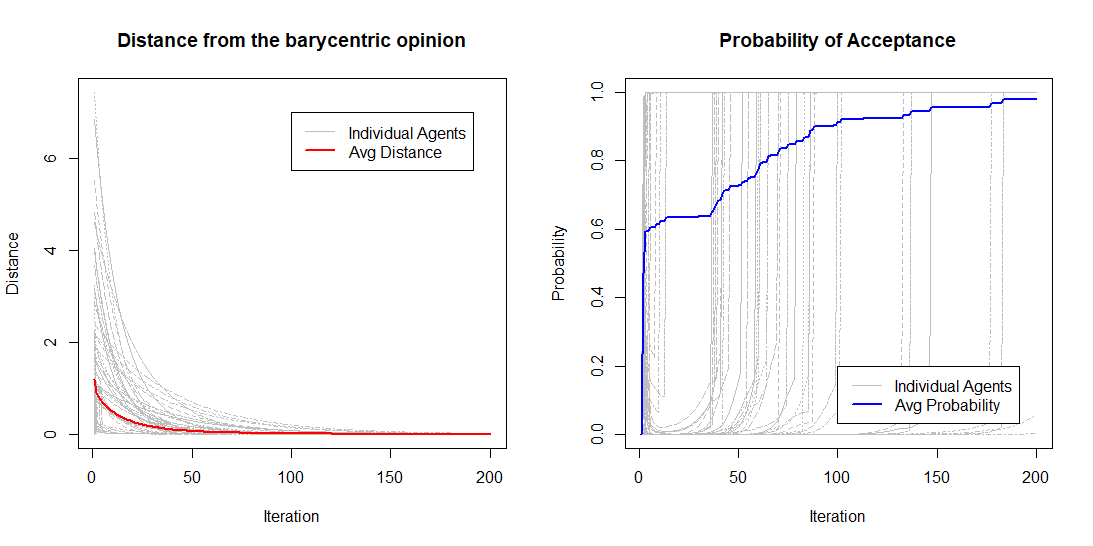}
	\caption{Convergence illustration to the barycenter by the one-stage process depicting for all agents: (a) distance from the consensus curve (left) and (b) acceptance probabilities with respect to the running barycentric curve}\label{fig-3}
\end{figure}	

In Figure \ref{fig-3} is illustrated a case of the one-stage scheme where a SDR-consensus is achieved for understanding the convergence of the scheme. On the left plot, each agent's divergence from the achieved consensus curve is illustrated. The red line, indicating the average distance of all agents from the consensus curve at each iteration, displays purely decreasing tendency. On the right plot, each agent's acceptance probability of the running consensus curve is illustrated, with the blue line indicating the average acceptance probability for all agents. It is also evident that the average acceptance probability displays purely increasing tendency to 1 as iteration number grows indicating converging behaviour to a consensus. In general, for any scenario considered, convergence is expected with potential differences in the convergence rates to be explained by the special characteristics of the scenario under study (different time-preferences of the involved agents). 

\begin{figure}[ht!]
	\centering
	\includegraphics[width=6in]{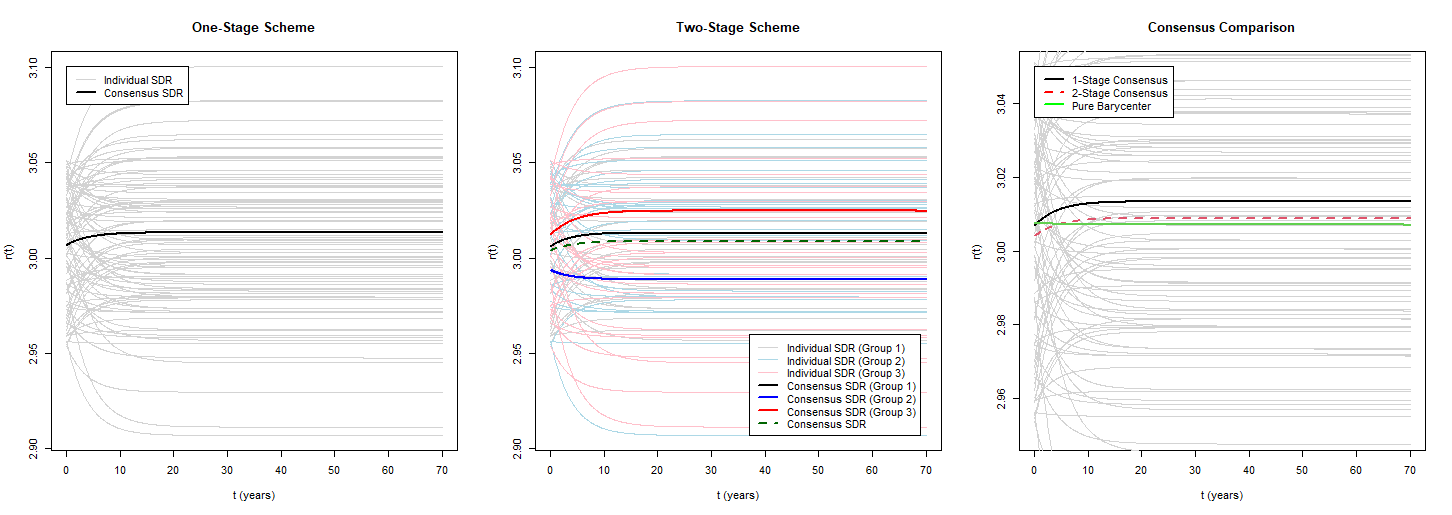}
	\includegraphics[width=6in]{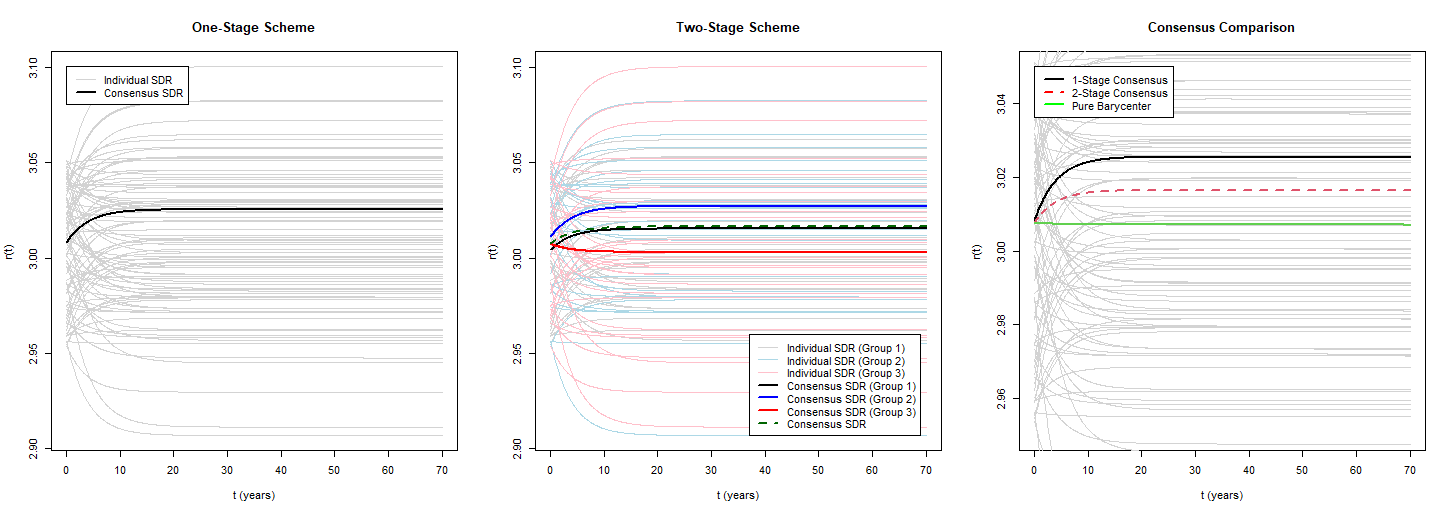}
	\caption{The achieved SDR-consensus curves achieved by the one-stage scheme (left), the two-stage process (center) and their comparison (right).}\label{fig-2}
\end{figure}	

In Figure \ref{fig-2} are illustrated the sampled SDR curves, their classification to the different subgroups (distinguished by different colours on the middle plot) and the obtained consensus curves by the two schemes for the Uniform Beliefs scenario (upper panel) and the Impatient Agents scenario (lower panel). For both scenarios are also illustrated the barycentric curves (no preferences taken into account) for comparison reasons. In both cases, the obtained consensus curve from the two-stage scheme seems to be less affected by the agents' preferences since it is closer to the pure barycentric curve than the one-stage consensus. However, both achieved consensus curves in both scenarios do not differ that much, and since the two-stage scheme is computationally cheaper should be preferred.

\begin{figure}[ht!]
	\centering
	\includegraphics[width=3in]{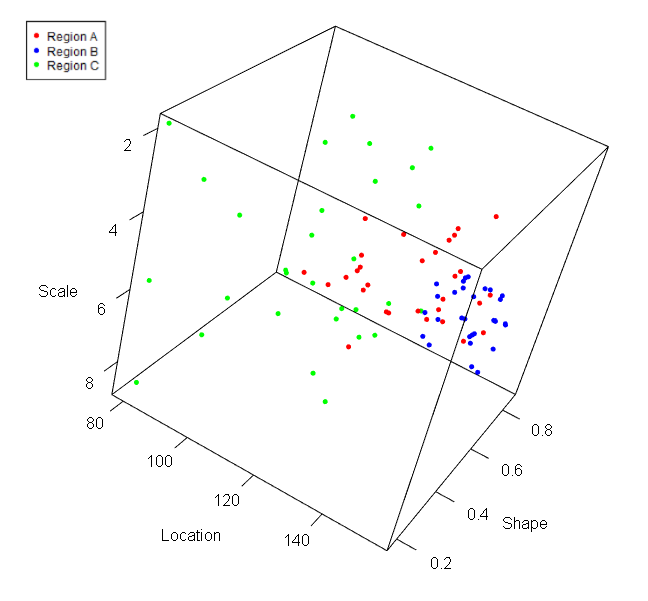}
	\caption{The simulated scenario for the beliefs concerning the probability model for the contingency}\label{fig-reg}
\end{figure}

As a second step, a consensus for the model describing the random behaviour (probability distribution) of the contingency $X$ at a future time $T$ is explored under both approaches and the two scenarios. Let us assume that all agents agree to the type of the model that could best describe the contingency distribution and in fact they consider the Generalized Extreme Value (GEV) distribution, which probability density function is 
\begin{equation}\label{400}
 f(x) = \frac{1}{\sigma} t(x)^{\xi + 1} e^{-t(x)},
 \end{equation}
with
\begin{equation*}
	t(x) = \left\{
	\begin{array}{ll}
		\left( 1 + \xi \left(\frac{x-\mu}{\sigma}\right) \right)^{-1/\xi}, & \mbox{ if } \xi \ne 0,\\
		e^{-(x-\mu)/\sigma}, & \mbox{ if } \xi=0,
	\end{array}\right.
\end{equation*}
where the parameters $\mu, \sigma>0, \xi$ capture the location, scale and shape characteristics, respectively. The difference in the agents beliefs are introduced through different estimates concerning the true parameter values. In particular we consider that within subgroups there is a short of homogeneity in the respective estimates (however not of the same level for all groups) while across the subgroups the heterogeneity level higher. An illustration of the scenario under consideration for the contingency probability model with respect to the parameter values is provided by Figure \ref{fig-reg}.

\begin{figure}[ht!]
\centering
\includegraphics[width=6in]{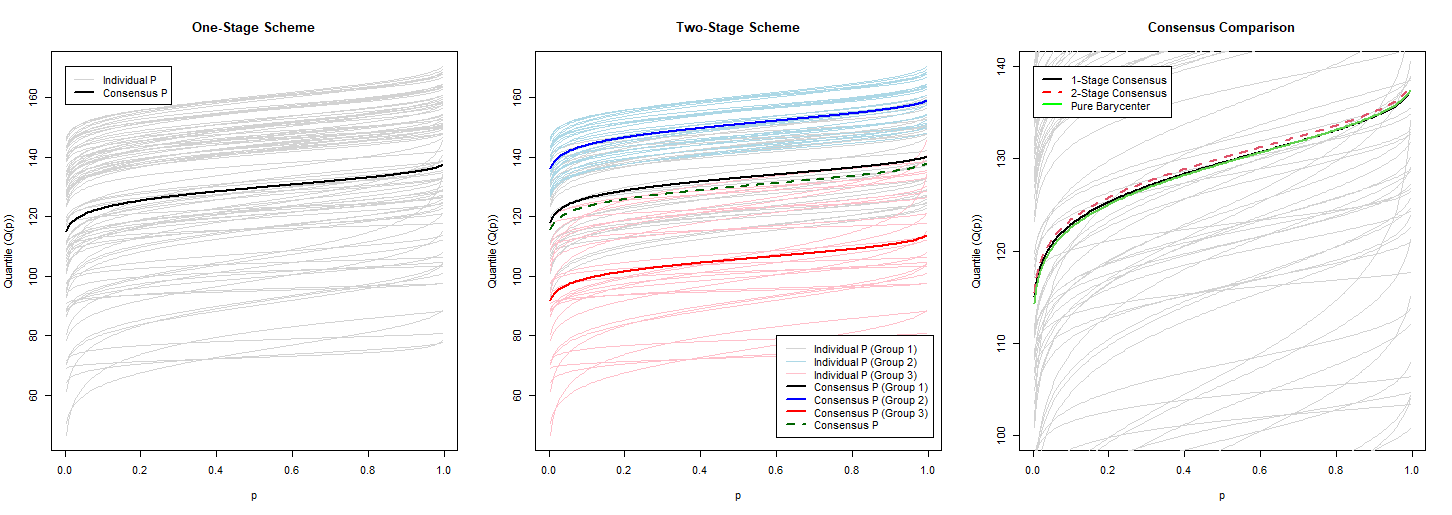}
\includegraphics[width=6in]{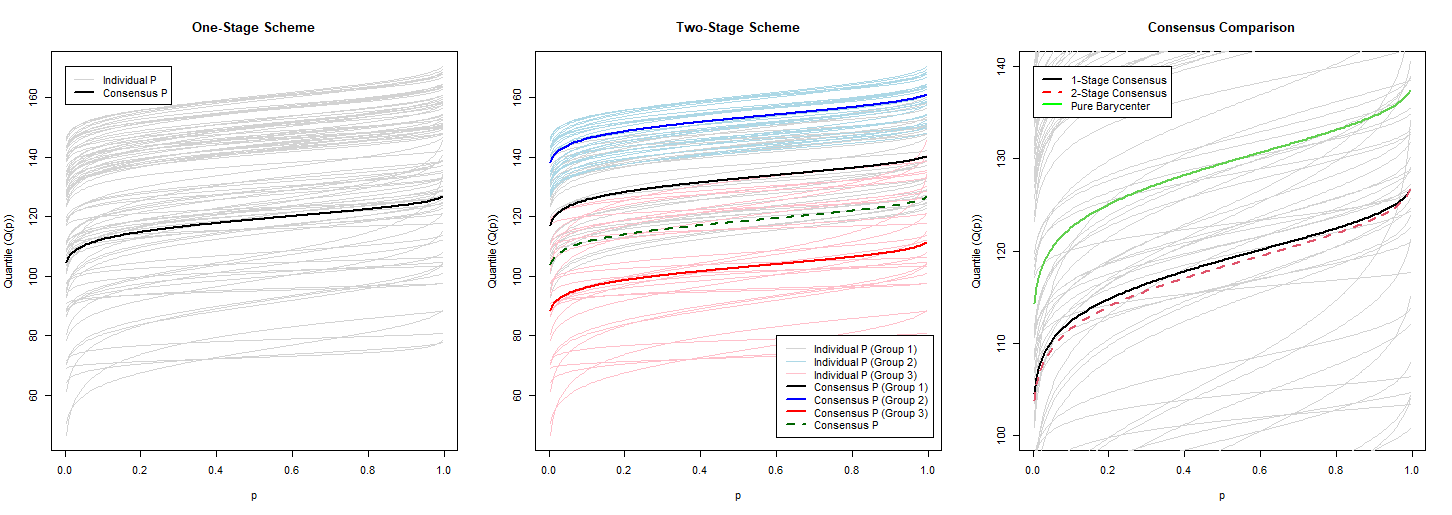}
	\caption{Achieved consensus from the one-stage scheme (left), the two-stage scheme (center) and their comparison (right) concerning the probability model that describes the contingency (in terms of quantiles) from the Uniform Beliefs scenario (upper panel) and the Impatient Agents scenario (lower panel).}\label{fig-5}
\end{figure}

Different considerations on the parameter vector $\theta = (\mu, \sigma\ \xi)'$ induce a different probability model $P$ describing the contingency $X$. As a result, the current set of opinions in this case is $\mathbb{M}=\{ P_1, ..., P_N \}$ which can be considered as a subset of the space of probability models in the real line, i.e. $\mathcal{M}=\mathcal{P}(\R)$. Since, this is the metric space (see Example \ref{WASSERSTEIN-METRIC-EX}) under which the consensus needs to be investigated, for the sake of simplicity, we assume that each provided $P_i$ is independent from the SDR curve $r_i(\cdot)$ provided by each agent. In Figure \ref{fig-5} are illustrated both scenarios and the achieved consensus models by the two schemes.

The consensus models obtained by both schemes for the two scenarios are quite close, however, the pure barycenter (direct quantile average in the initial beliefs) in the Impatient Agents scenario is quite far from the consensus indicating the effect of the agents' preferences in the derivation of the consensus. Combining the derived consensus opinions by both schemes, evaluation for the contingency under consideration is provided in Table \ref{tab-3} under the two scenarios, accompanied by some descriptive statistics to better quantify the differences in the estimation. The contingency evaluation is provided in present values discounted by the obtained SDR-curves by each scheme and the related consensus probability model.  Clearly, the estimates obtained in each scenario are quite close between the different approaches, however across the two scenarios, a significant difference is observed to the contingency valuation on account of the effect concerning different time-preferences of the involved agents.

\begin{table}[ht!]
	\centering
	\begin{tabular}{l|rr|rr}
		\hline\hline
		& \multicolumn{4}{|c}{\bf Scenario}\\
		\hline
	    {\bf Descriptive} & \multicolumn{2}{c}{\bf Uniform Beliefs} & \multicolumn{2}{|c}{\bf Impatient Agents}\\
	    {\bf Statistic}      & {\bf 1-Stage} & {\bf 2-Stage} &  {\bf 1-Stage} & {\bf 2-Stage}\\
	    \hline\hline 
Mean                      & 125.765  &125.260  &114.348   &114.971\\
Std. Deviation       &     4.349  &   4.344  &    4.425   &    4.284\\
1st-Percentile       &  114.578  &114.082  &103.126    &103.922\\
5th-Percentile      &117.923    &117.426   &106.446   &107.241\\
10th-Percentile    &119.856    &119.358   &108.369  &109.153\\
Median                  & 126.241   &125.737   &114.777   &115.443\\
90th-Percentile    &131.041    &130.531   &119.760   &120.162\\
95th-Percentile    &131.988   & 131.477   &120.794  &121.098\\
99th-Percentile    &133.198   & 132.686  &122.177   &122.305\\
		\hline\hline
	\end{tabular}
\caption{Descriptive statistics of the achieved consensus from the 1-Stage and the 2-Stage schemes for the contingency value for the two scenarios considered.}\label{tab-3}
\end{table}

\section{Conclusions}

In this paper we have considered the problem of group decision making under the effects of agents heterogeneity and model uncertainty. Our approach is  partly motivated by  situations commonly encountered in environmental economics, but the methodological framework has wider applicability. We propose an iterative procedure towards consensus, based on the concept of the Fr\'echet barycenter,  each step of which consists of  two stages: 
 the agents first update their position in the opinion metric space by a local barycenter characterized by the agents' immediate interactions and then a moderator makes a proposal in terms of a global barycenter, checking for consensus at each step. In cases of large inhomogeneous groups the procedure can be complemented by an auxiliary initial homogenization step, consisting of a clustering procedure in opinion space, leading to large homogeneous groups for which the aforementioned procedure will be applied.

 Our proposed evolutionary process towards consensus
 clarifies the effect of the behavioural characteristics of the agents on the effectiveness of the decision making process, the probability of reaching consensus and the expected time required for consensus.  The use of the method is illustrated by a characteristic problem of environmental economics, that of deciding on a common social discount factor and a common probabilistic model for future contingencies, which is to be used for pricing abatement measures and policy making in such a way as to be widely acceptable by the group, hence effective.

\appendix

\section{Proof of Proposition \ref{23-1}}\label{23-1-proof}

The proof uses a duality argument. To simplify the exposition we assume that $M$ is compact (or that we focus our attention on a compact subset of $M$). We define the function $L : M \times \R_{+} \times \R_{+}^{N+2} \to \R$ by
\begin{eqnarray*}
L(x,s,\mu_{00},\mu_0, \mu_1, \ldots, \mu_N) = s - \mu_{00} s + \mu_0( s -\epsilon_{min}^2) + \sum_{i=1}^{N} \mu_i (d^2(x,x_i)-s),
\end{eqnarray*}
and note that
\begin{equation}
\begin{aligned}
\max_{(\mu_{00}, \mu_0,\mu_1, \ldots, \mu_N) \in \R_{+}^{N+1}}& L(x,s,\mu_{00}, \mu_0, \mu_1, \ldots, \mu_N)= \\
& \left\{ \begin{array}{cccc} 
s & \mbox{if} & d^2(x,x_i) < s,  \,\, i=1, \ldots, N \,\,\,
   \& \,\,\, 0 \le s \le \epsilon_{min}^2, \\
+ \infty & & \mbox{otherwise}.
\end{array} \right .
\end{aligned}
\end{equation}
The variables $\mu_{00},\mu_0, \mu_1, \cdots, \mu_N$ play the role of Lagrange multipliers.

Hence problem \eqref{19-9-2023} can be written as
\begin{eqnarray}
\mbox{Problem} \eqref{19-9-2023}=\min_{x \in M, s \in \R_{+}} \max_{\mu \in \R_{+}^{N+2}}L(x,s,\mu),
\end{eqnarray}
where we used the compact notation $\mu=(\mu_{00}, \mu_0, \mu_1, \ldots, \mu_N)'$.
Using the minimax theorem  we can exchange the order of the maximum and the minimum to obtain
\begin{equation}\label{19-9-2023-1}
\begin{aligned}
\mbox{Problem} \eqref{19-9-2023}=\min_{x \in M, s \in \R_{+}} \max_{\mu \in \R_{+}^{N+2}} L(x,s,\mu)  \\= \max_{\mu \in \R_{+}^{N+2}}
\bigg( \underbrace{ \min_{x \in M, s \in \R_{+}} L(x,s,\mu)}_{:= D(\mu)} \bigg).
\end{aligned}
\end{equation}
The function
\begin{eqnarray}
D(\mu):=\min_{x \in M, s \in \R_{+}} L(x,s,\mu),
\end{eqnarray}
is called the dual function and as seen by \eqref{19-9-2023-1} can help characterize the solution of the primal problem \eqref{19-9-2023}.

We now proceed to the calculation of the dual function $D(\mu)$. By rearranging $L$ as
\begin{eqnarray}
L(x,s,\mu)=(1+\mu_0 - \mu_{00}- \sum_{i=1}^{N}\mu_i) s - \mu_0 \epsilon^2_{\min} + \sum_{i=1}^{N} \mu_i d^2(x,x_i),
\end{eqnarray}
we can see that the only interesting (non-generate) case corresponds to $\epsilon_1=\epsilon_2=...=\epsilon_N$ with $s>0$ (i.e. $\mu_{00}=0$) and $s=\epsilon_{\min}^2$ (i.e. $\mu_0, \mu_1,\mu_2,...,\mu_N>0$). This leads to the dual function
\begin{eqnarray}
D(\mu)=\min_{x \in M} \sum_{i=1}^{N} \mu_i d^2(x,x_i), \,\,\, \\
1 - \sum_{1}^{N}\mu_i =0.
\end{eqnarray}
In this case, the set of Lagrange multipliers $\mu_1,\mu_2,...,\mu_N$ can be realized as weights since they belong to $\Delta^{N-1}$. From now on let us denote $w=(\mu_1,\mu_2,...,\mu_N)'$ and express
\begin{eqnarray}
D(w)= \min_{x \in M} \sum_{i=1}^{N} w_{i} d^2(x,x_i).
\end{eqnarray}
The minimum over $x \in M$ is realized at the Fr\'echet barycenter of $\M$, with weight vector $w$ (as above). This implies that we should look for the solutions of problem \eqref{19-9-2023} among the Fr\'echet barycenters of $\M$. It remains to find the appropriate weights for the barycenter. According to the dual formulation \eqref{19-9-2023-1} the weights which are related to $\mu>0$ are obtained as the solution of the dual problem
\begin{eqnarray*}
\max_{\mu \in \R_{+}^{N+2}} D(\mu) =\max_{w \in \Delta^{N-1}} V_{\M} (w),
\end{eqnarray*}
where $V_{\M}(w)$ is the Fr\'echet variance of $\M$ when choosing $w$ as the weight vector. Hence we conclude that the solution of \eqref{19-9-2023} is a Fr\'echet barycenter with a weight chosen so as to maximize the corresponding weighted Fr\'echet variance.

\section{Proof of Proposition \ref{23-2}}\label{23-2-proof}

The proof proceeds in several steps.

\emph{Step 1}. We note that problem \eqref{max-agree-prob} is equivalent to the minimization problem
\begin{eqnarray}\label{min-agree-prob}
\min_{x \in M} \sum_{i=1}^{N} \psi_{i}(d^2(x,x_{i})),
\end{eqnarray}
where $\psi_i=-\ln \phi_i$. This is easy to see, as maximizing $P$ is equivalent to minimizing $-\ln P$. Note that the functions $\psi_i$ are increasing.

\emph{Step 2}. Let $\partial \phi$ be the subdifferential of a  (convex) function $\psi$. The first order condition for $\bar{x}$ to be a local minimizer of $\psi$ is that $0 \in \partial \psi(\bar{x})$. Applying Assumption \ref{ASS-1}, we assume that for each $i=1, \ldots, N$ we can express $x_i=x_i(\theta_i)$ for some $\theta_i \in H$, and similarly, any point $x \in X$ can be expressed as $x=x(\theta)$ for some $\theta \in H$. To simplify notation we will not revert to the $\theta$ parametric notation but keep our initial notation in terms of $x$. We apply the first order condition to $\psi(x):=\sum_{i=1}^{N}\psi_i(d^2(x,x_i))$, which in  parametric form becomes
$$\psi(\x):=\sum_{i=1}^{N} \psi_i(d^2(x(\x),x_i(\x_i))=\sum_{i=1}^{N} \psi_i(d^2(\x,\x_i)),$$ with the latter used as a simplification of the notation.  The problem of minimizing over $x \in X$ is then transferred to minimizing over the parameter $\x \in H$. This yields
\begin{eqnarray}\label{1234}
0 \in \partial \sum_{i=1}^{N} \psi_i(d^2(\x,\x_i)) = \sum_{i=1}^{N} \partial \psi_i(d^2(\x,\x_i)),
\end{eqnarray}
where for the second equality we have used the subdifferential calculus (which holds since $\psi_i$ and $d^2(x,x_i)$ are continuous. We now apply the subdifferential rule for composite functions (see e.g. Corollary 6.72 in \cite{bauschke2017correction}. This yields that for each $i=1,\ldots, N$ it holds that $z \in \partial \psi_i(d^2(\x,x_i))(y) $ if there exists $\alpha_i \in \partial \psi_i (d^2(y,x_i))$ and $w_i \in \partial d^2(\cdot, x_i) (y)$ such that $z=\alpha_i w_i$. To simplify the exposition let us assume that $\psi_i$ is $C^1$, so that the subdifferential of $\psi_i$ is a singleton consisting of a single value, that of the derivative $\psi_i'(d^2(y,x_i)) >0$, with its positivity guaranteed by the fact that $\psi$ is increasing. Condition \eqref{1234} together with the above subdifferential rule implies that
\begin{eqnarray}\label{2345}
0 \in \sum_{i=1}^{N} \bar{w}_{i} \partial d^2(\x,\x_i)(\x^*),
\end{eqnarray}
where $\bar{w}_i=\psi_i'(d^2(\cdot, x_i))(\x^*) >0$. Defining $w_{i}=\frac{\bar{w}_i}{\sum_{i=k}^{N} \bar{w}_{k}} > 0$, and dividing \eqref{2345} by $\sum_{k=1}^{N} \bar{w}_{k}$ we obtain 
\begin{eqnarray}\label{3456}
0 \in \sum_{i=1}^{N} w_{i} \partial d^2(\x,\x_i)(\x^*),
\end{eqnarray}
where $w=(w_1, \ldots, w_N)\in \Delta^{N-1}$. From \eqref{3456} we conclude that $\x^*$ is the solution of the minimization problem
\begin{eqnarray}
\min_{\x \in H} \sum_{i=1}^{N} w_i d^2(\x,\x_i),
\end{eqnarray}
i.e. $x^*=x(\x^*)$ corresponds to a Fr\'echet barycenter  for the choice of weights $w=(w_1, \ldots, w_N) \in \Delta^{N-1}$.

\emph{Step 3}. It remains to show that such a choice of weights is feasible. The weights $w=(w_1, \ldots, w_N)$ must be such that if $x^*=x(\x^*)$ is the solution of \eqref{3456} then  it must hold that
\begin{eqnarray}\label{4567}
w_i =\psi_i'(d^2(x^*,x_i))=\psi_i'(d^2(\x^*,\x_i)), \,\,\, i=1,\ldots, N.
\end{eqnarray}
Since $x^*=x(\x^*)$ is a function of the weights vector $w \in \Delta^{N-1}$, we can interpret \eqref{4567} as a system of equations for $w$, the solution of which in $\Delta^{N-1}$ will characterize the appropriate weights for the barycenter at which consensus can be reached. Defining the function $w \to g(w):=x(\x^*)$ where $\x^*$ is the solution of problem \eqref{3456}, we can rewrite \eqref{4567} as
\begin{eqnarray}\label{5678}
w_{i}=\psi_i'(d^2(g(w), x_i)), \,\,\,\, i=1, \ldots, N.
\end{eqnarray}
If the functions $w \mapsto \psi_i'(d^2(g(w),x_i))$ are continuous then a solution to \eqref{5678} in $\Delta^{N-1}$ is guaranteed by Brouwer's fixed point theorem. It thus remains to show the continuity of the functions $w \mapsto \psi_i'(d^2(g(w),x_i))$. Since $\psi_i'$ are continuous it suffices to show the continuity of the function $w \to g(w):=x(\x^*)$ where $\x^*$ is the solution of problem \eqref{3456}.

\emph{Step 4.} We will use the following definitions: Let $(w^{(n)})$ be a sequence in $\Delta^{N-1}$ and define the sequence of functionals $\psi^{(n)} : M \to \R$, by 
\begin{eqnarray*}
\psi^{(n)}(x)=\sum_{i=1}^{N} w^{(n)}_i d^2(x, x_i).
\end{eqnarray*}
If needed (which is not required here) we may express this in terms of the parametric representation $x=x(\x)$, as a functional on the space of parameters $H$.

\emph{Step 5.} We claim the following:  Consider a sequence $(w^{(n)}) \subset \Delta^{N-1}$, such that $w^{(n)} \to w$ in $\Delta^{N-1}$, and a sequence $(x^{(n)}) \subset M$ such that $x^{(n)} \to x$ in $M$. Then, $\psi^{(n)}(x^{(n)}) \to \psi(x)$ in $\R$, where $\psi(x)=\sum_{i=1}^{N} w_{i} d^2(x,x_i)$. To see that, we calculate
\begin{eqnarray*}
| \psi^{(n)}(x^{(n)}) -\psi(x) | = | \sum_{i=1}^{N} w_{i}^{(n)} d^2(x^{(n)},x_i) - \sum_{i=1}^{N} w_i d^2(x,x_i) |\\
=| \sum_{i=1}^{N} (w^{(n)}_i -w_i) d^2(x^{(n)},x_i) + \sum_{i=1}^{N} w_{i} (d^2(x^{(n)},x_i)-d^2(x,x_i) ) | \\
\le \sum_{i=1}^{N} | w^{(n)}_i -w_i| d^2(x^{(n)},x_i) + \sum_{i=1}^{N} |d^2(x^{(n)},x_i)-d^2(x,x_i)| \to 0,
\end{eqnarray*}
with the first term tending to 0 since $w^{(n)}\to w$ and the second term tending to $0$ since $x^{(n)} \to x$.

\emph{Step 6.} We also claim the following: Consider the  sequence $(x_{*}^{(n)}) \subset M$ of minimizers of the sequence of functionals $(\psi^{(n)})$  (i.e. for every $n \in \N$, $x^{(n)}=\arg\min_{x} \psi^{(n)}(x)$).  Then, if $x^{(n)} \to x$ in $M$, it follows that $x =\arg\min_{x \in M} \psi(x)$. To show this, consider any $z \in M$ and a sequence $(z^{(n)}) \subset M$ such that $z^{(n)}\to z$ in $M$. Then,
\begin{eqnarray*}
\psi(x)\stackrel{\mbox{by step} \,\, 5}{=} \lim_{n} \psi^{(n)}(x^{(n)}) =
\lim\inf_{n} \psi^{(n)}(x^{(n)}) \stackrel{ \psi^{(n)}(x^{(n)}) \le \psi^{(n)}(z^{(n)})}{\le} \\
\lim\inf_{n} \psi^{(n)}(z^{(n)}) \le \lim\sup_{n} \psi^{(n)}(z^{(n)})
\stackrel{by step  5}{=} \psi(z),
\end{eqnarray*}
hence, by the fact that $z$ is arbitrary, $x = \arg\min_{x \in M} \psi(x)$.

\emph{Step 7.} We are now in position to show the continuity of the barycenter with respect to the weights, i.e. the continuity of the map  $w \to g(w):=x(\x^*)$ where $\x^*$ is the solution of problem \eqref{3456}. Let $(w^{(n)} \subset \Delta^{N-1}$ be a sequence of weights and  for every $n \in \N$, let $x_{*}^{(n)}$ be the barycenter of $\M$ for the weight vector $w^{(n)}$, i.e. the minimizer for the functional $\psi^{(n)}$. Assume that the sequence $(x^{(n)})$ (or a subsequence) has a limit $x \in M$. This can be achieved by a compactness argument (or weak compactness). Then,  by step 6, $x$ is the minimizer of $\psi$, hence the barycenter for the weight vector $w$. Hence, the map $g$ is continuous.

This concludes the proof.

\section{Proof of the claim in Example \ref{NORMAL-BARY} }\label{NORMAL-BARY-PROOF}

We recall (see e.g. \cite{bhatia2019bures})  that between two measures $P_i \sim N(\mu_i, S_{i})$, $i=1,2$, the Wasserstein distance $W_{2}^{2}(P_1,P_2)$, admits the closed form 
\begin{eqnarray}\label{APP-200}
W_2^2(P_1,P_2)=\| \mu_1 -\mu_2\|^2 + Tr\bigg(S_1 + S_2 - 2 ( S_1^{1/2} S_2 S_1^{1/2})^{1/2}\bigg).
\end{eqnarray}
Moreover, given a set of probability measures ${\mathbb M}$ consisting of Gaussian measures $P_i$, $i=1, \ldots, M$, and a weight vector $(w_1, \ldots, w_{K})$, the corresponding Wasserstein barycenter $P_{B}$ is a Gaussian measure $P_{B} \sim N(\mu, S)$ with $\mu= \sum_{i=1}^{K} w_{i} \mu_{i}$, and $S$ being a matrix that satisfies the equation
\begin{eqnarray}\label{LIN}
    0=I-\sum_{k=1}^{K} w_{k} (S_{k} \# S^{-1}) \,\, \Longleftrightarrow \,\, S = \sum_{k=1}^{K} w_{k} (S^{1/2} S_{k} S^{1/2})^{1/2},
\end{eqnarray}
where the notation $A \# B$ is used to denote the geometric mean between two positive definite symmetric matrices given by
\begin{eqnarray*}
    A \# B = A^{1/2}(A^{-1/2} B A^{-1/2})^{1/2} A^{1/2} = B \# A.
\end{eqnarray*}
Without loss of generality we will assume that $\mu_k =0$, $k=1,\ldots, K$ (else simply center the measures). We will also consider problem \eqref{max-agree-prob} on ${\cal N}(\R^{d}) \subset {\cal P}(\R^{d})$, the subset of Gaussian measures on $\R^d$. With the above information problem \eqref{max-agree-prob} can be expressed as
\begin{eqnarray}\label{MIN-B}
    \max_{S} \Psi(S):= \max_{S} \sum_{k=1}^{K} \psi_{k} \bigg( Tr(S_k) + Tr( S - 2 g_{k}(S)) \bigg),
\end{eqnarray}
where $\psi_k =\ln \phi_k$ and $g_{k}(S):= (S_{k}^{1/2}S S_{k}^{1/2})^{1/2}$. Problem \eqref{MIN-B} is an optimization problem  on the set of positive definite symmetric matrices. It can be treated by considering the Fr\'echet derivative of the functional in \eqref{MIN-B} with respect to $S$. Using the rules of Fr\'echet differentiation and assuming sufficient smoothness for the functions $\psi_{k}$ we have that for any deviation $S+ \epsilon Z$ from the matrix $S$ the action of the Fr\'echet derivative $D\Psi(S)$ on any matrix $Z$ yields
\begin{eqnarray}\label{AAA-BBB}
    [D\Psi(S)]Z = \sum_{k=1}^{K} \psi_{k}'(W_{k}) Tr(Z - [Dg_{k}(S)]Z),
\end{eqnarray}
where we use the simplified notation
\begin{eqnarray*}
    W_{k}= Tr(S_k) + Tr( S - 2 g_{k}(S)).
\end{eqnarray*}
Moreover, define the quantities
\begin{eqnarray*}
    \Lambda_{k} = \psi_{k}'(W_{k}) \in \R_{+},
\end{eqnarray*}
where the positivity of $\Lambda_k$ is guaranteed by the properties of the functions $\psi_{k}$. Following \cite{bhatia2019bures}, we can compute
\begin{eqnarray*}
Tr (Dg_{k}(S) Z) = Tr( (S_{k} \# S^{-1}) Z ),
\end{eqnarray*}
so that \eqref{AAA-BBB} yields (using the linearity of trace) that
\begin{eqnarray*}
    [D\Psi(S)]Z =Tr[ (\sum_{k=1}^{K} \Lambda_k) I - \sum_{k=1}^{K} \Lambda_{k}  (S_{k} \# S^{-1}) ) Z].
\end{eqnarray*}
The first order condition for the solution of \eqref{MIN-B} is $[D\Psi(S)]Z=0$, for all possible perturbations $Z$  of the covariance matrix $S$. Upon defining
\begin{eqnarray*}
w_{k} = \frac{\Lambda_{k}}{\sum_{j=1}^{K} \Lambda_{j}} \in [0,1], \,\,\, k=1, \ldots, K,
    \end{eqnarray*}
the first order condition becomes
\begin{eqnarray*}
    Tr[ (I -\sum_{k=1}^{K} w_{k} (S_{k} \# S^{-1}) ) Z ]=0, \,\,\, \forall \, Z,
\end{eqnarray*}
which implies that the solution of\eqref{MIN-B} corresponds to a Gaussian measure with covariance matrix $S$ such that
\begin{eqnarray}\label{NONL}
I -\sum_{k=1}^{K} w_{k} (S_{k} \# S^{-1})  =0 \,\, \Longleftrightarrow \,\, S = \sum_{k=1}^{K} w_{k} (S^{1/2} S_{k} S^{1/2})^{1/2}, 
\end{eqnarray}
i.e. $P^{*}$ is the barycenter of ${\mathbb M}$ with a selection of weights $w_{k}$, endogenously obtained by the preferences on the agents towards their anchor point (in other words their bargaining power). 

Note that equation \eqref{NONL}, although formally the same as equation \eqref{LIN} has a fundamental difference from \eqref{LIN}. In \eqref{NONL} the coefficients $w_{k}=w_{k}(S)$, i.e. are depending on $S$, whereas in \eqref{LIN} the coefficients $w_{k}$ are constants. It remains to show that equation \eqref{NONL} admits a solution. To show that we define the operator $T$, by $S \mapsto T(S):=\sum_{k=1}^{K} w_{k} (S^{1/2} S_{k} S^{1/2})^{1/2}$. It can be shown that this operator maps the closed convex set ${\cal K}=\{ S \in \R_{+}^{d \times d} \,\, \mid \,\, c_1 I \le S \le c_2 I\} $, where $c_1,c_2 \ge 0$ and  by $\le$ we denote the natural ordering $S_1 \le S_2 \,\, \Longleftrightarrow S_1 -S_2 \ge 0$ (meaning $S_1-S_2$ positive definite) onto itself. The set ${\cal K}$ is convex, and the map $T$ is continuous, so by the Brouwer fixed point theorem $T$ has a fixed point, therefore \eqref{NONL} admits a solution.

\section{Extensions to the evolutionary algorithm}\label{EXTENSIONS}

The evolutionary scheme presented in this section can be further extended.

\subsection{A two stage scheme involving a clustering step for opinion homogenization and group formation}\label{DCP-SEC}

As indicated by the numerical experiments in Section \ref{sec-3.2} large degree of inhomogeneity of the positions of the agents in opinion space, especially in cases of large groups,  may lead to delay in the convergence to consensus. In certain cases (i.e. in cases of emergency etc) such delays may be unwanted. A way to avoid situations like this is to find ways of grouping the $N$ agents in $K$ as far as possible homogeneous subgroups, \footnote{For instance, the large group could be the general population of a country, whereas the clusters may correspond to tendencies within the country. As another example we may consider a group of $N$ consumers which is further clustered into $K$ homogeneous subgroups in terms of preferences.} each one characterized  by a representative opinion $\bar{x}_{k}$, $k=1,\ldots, K$, and then performing the consensus procedure described in Section \ref{23-9-100} among them.

To this end we propose the following version of the celebrated K-means clustering algorithm. We will consider the opinions of the large groups as elements $\{x_1, \ldots, x_{N}\}$  of the opinion metric space $(M,d)$. The idea is that like opinions will form clusters in this metric space. Upon being able to identify these clusters we can form a coarse graining of the group into sub-groups of like opinions, which can be treated as homogeneous groups for our level of coarse graining.  Mathematically, this corresponds to breaking the large group $G$ into $k$ subgroups $G_{i}$, $i=1,\ldots, k$, such that $G = \bigcup_{i=1}^{k} G_{i}$, and $G_{i} \cap G_{j} = \emptyset$  for $i \ne j$, with the opinions $x_{\ell } \in G_{i}$ being as homogeneous as possible. As discussed above, homogeneity of a subgroup will be understood in terms of the Fr\'echet function of the subgroup, whereas a relevant measure for the center of the group will be the Fr\'echet barycenter of the subgroup. This scheme can be applied for any relevant metrization of the opinion space $M$ (see e.g. examples in previous section), for the case of the Wasserstein space see  \cite{papayiannis2021clustering}. The proposed clustering algorithm to be implemented in the opinion space is summarized in Algorithm \ref{alg-1}.

\begin{algorithm}\caption{K-Means Clustering Scheme in the Opinions Space}\label{alg-1}
\begin{enumerate}
\item Choose a relevant metrization of the opinion space $(M,d)$ and a number of clusters $K$ with centers $\bar{x}_{k}^{(0)}$ for $k=1, \ldots, K$.
\item At each step $m$, each of the opinions $x_{i}$ for $i=1, \ldots, N$ is assigned to one of the $K$ clusters where the cluster membership $k(i) \in \{1, \ldots, K\}$ is determined according to the rule
\begin{eqnarray*}
k(i) \in \arg \min_{k \in \{1, \ldots, K\} } d(x_{i}, \bar{x}_{k}^{(m)}).
\end{eqnarray*}
\item Cluster centers are updated through the rule
\begin{eqnarray*}
 \bar{x}_{k}^{(m+1)}=\arg \min_{ z \in M} \frac{1}{n_{k}^{(m)}} \sum_{i=1}^{n_k^{(m)}} d^2(x, x_{k,i}^{(m)}), \,\,\, k=1, \ldots, K,
\end{eqnarray*}
where $n_{k}^{(m)}$ is the number of points that have been assigned to cluster $k$ and by $x_{k,i}^{(m)}$, $i=1, \ldots, n_{k}^{(m)}$ we denote the points  that have been assigned to cluster $k$, at step $m$ of the algorithm.
\item Steps 2-3 are repeated until the cluster centers do not change significantly.
\end{enumerate}
\end{algorithm}

At the convergence of the algorithm, $K$ clusters of opinions are determined, centered at the points $\bar{x}_{k}$, $k=1, \ldots, K$ in opinion space $(M,d)$. Each of these clusters can be understood as a more or less ``homogeneous'' group of agents in terms of opinions. Denoting the groups by $G_{k}$, $k=1, \ldots, K$, we expect our clustering algorithm to perform well in segregating the general group of agents $G$ into subgroups if the Fr\'echet variance of each subgroup $V_{k}:= \min_{z \in M} F_{G_{k}}(z)$ is comparatively low. Recall that the Fr\'echet variance of a subset $G_{k} \subset M$ can be also understood as an indicator of its homogeneity.

Note that the above algorithm can be expressed in terms of an optimization problem of the form
\begin{eqnarray}
\min_{\substack{ \bar{x}_{k} \in M, \\  k=1, \ldots, K} } \sum_{j=1}^{K} \sum_{i=1}^{N} a_{ij} d^2(x_{i}, \bar{x}_{k}),     \label{k-means-min} \\
\mbox{where} \,\,\, a_{ij}=\left\{ \begin{array}{ccc} 1  & \mbox{if} & j=\arg\min_{\ell} d(x_{i},\bar{x}_{\ell}), \\
0 &  &\mbox{otherwise}.
\end{array}\right. \label{k-means-a} 
\end{eqnarray}
In other words, the elements $a_{ij}$ provide information as to the membership of the point $i$ to the cluster $j$, taking the value 1 if $i$ belongs to cluster $j$ and $0$ otherwise. The $K$-means algorithm solves this problem by the following two-step procedure iterating Steps A and B till convergence:
\begin{itemize}
\item[A.] Given the centers $\bar{x}_{j}$, calculate $a_{ij}$ solving the minimization problem  \eqref{k-means-a}. This generates a membership matrix $A=(a_{ij}) \in \R^{N \times K}$ containing binary entries, with each column $k$ of $A$ denoting the composition of the group $G_{k}$.
\item[B.] Given the solution for $a_{ij}$ from Step A, the new centers are determined by solving \eqref{k-means-min}. Note that this step breaks down into $K$ decoupled problems, each one involving the minimization of the Fr\'echet function for each $G_{k}$, or equivalently finding the Fr\'echet mean of the group, which is recognized as the center of the corresponding cluster. The objective's value at the minimum will then be the sum of the Fr\'echet variances of the clusters $\sum_{k=1}^{K} V_{G_{k}}$.
\end{itemize}

We close this section by summing up the two-stage group decision process to the following steps:

\begin{enumerate}
    \item Collect and map all opinions of the group  as points $z_i$, $i \in G$ into the appropriate opinion space $\mathcal{M}$. 
    \item Perform a clustering procedure in opinion space as proposed in Algorithm \ref{alg-1} to form $K$ groups in opinion space. 
    \item Identify the group agreement point as a barycenter of the set of opinions ${\mathbb M}=\{z_{B,1}, \ldots, z_{B,K}\}$, i.e. as a barycenter of barycenters. 
\end{enumerate}

\subsection{Updating the discounting parameter}

Possible extensions of the scheme where the sensitivity parameter $r_i$ could be time-varying can be conceived. For example  a possible evolution scheme for the parameter could be described as
\begin{equation}
    r_i(t+1) = L_i(t, \mathbb{M}(t)),\,\,\,\, \forall i\in G,
\end{equation}
considering $L_i$ as the loss function of the agent $i$ depending on the states of the current opinion set $\mathbb{M}(t)$, i.e. indicating the loss (under the assumption that $L_i\geq 0$) taking into account the time $t$ and the level of homogeneity in the opinion set $\mathbb{M}(t)$. For instance, a possible choice could be the subjective rule
\begin{equation}\label{time-update}
r_i(t+1) = \left\{
\begin{array}{ll}
0, & \mbox{if } d^2(\mu_i(0), \mu_j(t)) \leq \epsilon_i, \forall j,\\
e^{-\rho_i t} \sum_{j\in\mathcal{N}_i} w_{ij}(t) d^2(\mu_i(0), \mu_j(t)), & \mbox{if } d^2(\mu_i(0), \mu_j(t)) > \epsilon_i, \forall j,
\end{array} \right.
\end{equation}
where $\rho_i > 0$ expresses the agent's preferences concerning a fast resolution of the problem while $\epsilon_i>0$ denotes the agents desire to deviate from her/his anchor preferences $\mu_i(0)$. In this setting, the preferences concerning the time upon which a consensus should be reached governs the determination of the general time-preferences parameter as $t$ grows.  

\section{On the choice of metrics and the calculation of the Fr\'echet mean}

\subsection{Choice of metric and the Fr\'echet mean} 

The choice of metric depends on the opinion space. For instance, in the Example \ref{METRIC-CURVE} concerning the social discount rate and the valuation of future uncertain costs, a possible choice of metric space could be determined in thew following way. There are two important components of the opinion space (a) the yield curve characterizing the discount rate and (b) the probability distribution of the future risk to be evaluated. Then, the natural choice of the opinion space is a Cartesian product of two metric spaces $M=M_1 \times M_2$, where $M_1$ corresponds to the metric space of possible yield curves and $M_2$ corresponds to the metric space of possible probability models for the future risk. Let us consider each one separately.

\subsubsection*{The space of discount rate curves}

As we denoted, $M_1$ is a space of curves $f : [0,T] \to \R_{+}$, where $T$ is the time horizon in question. Not any curve is a representation of a suitable term structure model, so we restrict ourselves  to parametric families of curves, that are consistent with economic reasoning. One possible choice for $M_1$ could be the parametric family of curves ${\cal R}$. A suitable example can be the set defined in \eqref{R-set}. To simplify the  notation, we will denote all relevant parameters as $\theta \in \Theta$, where $\Theta$ is a suitable parameter space (e.g. $(y_1,\phi)$ in the case of ${\cal R}$ in \eqref{R-set}). We will denote any element of $M_1$ as $\Phi(\cdot ; \theta)$, meaning a function $t \mapsto f(t)=\Phi(t; \theta)$, parameterized by some $\theta \in \Theta$ and a suitable function $\Phi : [0,T]\times \Theta \to \R$. By definition we can identify any element $f \in M_1={\cal R} \subset L^{2}(0,T)$ by an element $\theta \in \Theta \subset \R^{d}$. We will denote by $\Phi^{-1}$ the mapping ${\cal R} \to \Theta$, that assigns to any $f \in {\cal R}$ a relevant $\theta \in \Theta$, such that $f(\cdot)=\Phi(\cdot ; \theta)$. We do not necessarily require this mapping to be single valued (but we require $\Phi$ to satisfy this property).

As stated above, $M_1={\cal R}$ is not a linear space, since linear combinations of functions $f \in {\cal R}$ are not necessarily elements of ${\cal R}$. There are various choices for the metric on $M_1$.

One possible choice of metric on $M_1={\cal R}$ could be in terms of a suitable metric for $\Theta \subset \R^{d}$, i.e.
\begin{eqnarray}\label{Theta-Metric}
d_{M_1}(f_1, f_2)= d_{\Theta}(\Phi^{-1}(f_1), \Phi^{-1}(f_2)) = d_{\Theta}(\theta_1, \theta_2),
\end{eqnarray}
where $\theta_i$ are such that $f_{i}(\cdot)=\Phi(\cdot ; \theta_i)$, $i=1,2$, and $d_{\Theta}$ is a suitable metric for $\Theta \subset \R^{d}$ (a possible choice being an $\ell_{p}$ metric, e.g. the Euclidean metric).  The nonlinear nature of the transformation $\Phi$, turns $d_{M_1}$ into a metric on $M_1$ which is not directly  related to a metric derivable from a norm on $L^{2}(0,T)$, eventhough it may be related to a norm in the parameter space $\Theta$.

Another possible choice could be a metric compatible with the vector space in which the nonlinear set ${\cal R}$ is naturally embedded, a choice for which could be the Hilbert space $L^{2}(0,T)$. Hence a suitable metric for $M_1$ could be as follows: For any two $f,f' \in M_1= {\cal R}$ there exists a pair $\theta, \theta' \in \Theta$ such that $f$ can be identified by $t \mapsto \Phi(t ; \theta)$ and $f'$ can be identified as $t \mapsto \Phi(t ; \theta')$. Then we may define the metric
\begin{eqnarray}\label{APP-100}
d_{M_1}(f,f')= \bigg( \int_{0}^{T} |\Phi(t, \theta) -\Phi(t, \theta')|^2 dt  \bigg)^{1/2} =: \hat{d}(\theta, \theta').
\end{eqnarray}
Note that while this metric formally coincides with the $L^2$ norm, the space $(M_1, d)$ is not a vector space on account of the nonlinearity of $M_1$, eventhough it is endowed with a metric compatible with the norm of the vector space in which $M_1$ is embedded in. This metric is typically used for spaces of curves of a given parametric representation, i.e. for spaces of curves of the general form ${\cal R}:=\{ t \mapsto \Phi(t; \theta) \,\, : \,\, t \in [0,T], \,\, \theta \in \Theta\}$, as for example in the shape invariant model which is commonly used in functional data analysis (see e.g. \cite{bigot2013frechet}, or
\cite{papayiannis2023modelling} and references therein). Other choices are of course possible, for alternative choices  we refer the reader to \cite{srivastava2016functional}, where a detailed discussion of related issues is provided, see also \cite{steyer2023elastic} for recent applications.

For the numerical illustrations presented in this paper, we opted for the choice \eqref{Theta-Metric}, rather than \eqref{APP-100}, mainly for two reasons: (a) Since often decision making concerning scientific issues is model based, consensus upon a  model is naturally reduced to consensus on the parameters of the model and (b) the consensus path in the finite parameter space is easier to illustrate and visualize. On the other hand, experiments were also performed using the alternative metric \eqref{APP-100}, and the results were qualitatively and quantitatively similar. 

For any of the above choices, the Fr\'echet mean in $M_1$ is now defined as follows: 
\begin{eqnarray}\label{APP-101}
f_{B}=\arg\min_{f \in M_1} \bigg( \sum_{i=1}^{N} w_i d^2(f, f_i ) \bigg) 
=\arg\min_{\theta \in \Theta} \sum_{i=1}^{N}w_{i} \hat{d}^2(\theta, \theta_i),
\end{eqnarray}
where $\hat{d}$ is defined as in \eqref{APP-100}.
It is important to note that as defined, $f_{B} \in {\cal R}$, and is not merely an element of the embedding space of ${\cal R}$, which is $L^2(0,T)$. Moreover, we must stress that the averaging obtained in \eqref{APP-101} is a nonlinear averaging of the yield curves $f_{i} \in {\cal R}$, and not the standard linear averaging $f_{A}=\sum_{i=1}^{N} w_{i} f_{i}$ in the embedding space $L^{2}(0,T)$, with $f_{A}$ typically having the property $f_{A} \not\in {\cal R}$, hence being not acceptable as a suitable yield curve.

\subsubsection*{The space of probability models for the future risk}

We now consider the space of possible distributions of future risks. This a space of probability distributions on $\R^d$, so a suitable choice of opinion space would be $M_2= {\cal P}(\R^d)$, the space of probability measures on $\R^d$. This is again a space that does not admit a vector space structure. A suitable metrization is in terms of the Wasserstein metric presented in Example \ref{WASSERSTEIN-METRIC-EX}. In the case where the risks can be represented by a single random variable (i.e. in terms of their pecuniar value only) we can consider distributions on $\R$. These can be represented as distribution functions $F$, hence $M_2$ can be identified with the space of distribution functions. This is clearly not a vector space, but may be embedded on a suitable vector space of functions (e.g. measurable functions).  As it turns out, it is not the actual distribution function which is important in the metrization of ${\cal P}(\R)$, but rather its generalized inverse, the quantile function $Q:=F^{-1}$.  For $M_2={\cal P}(\R)$, the Wasserstein metric can be conveniently expressed in terms of quantiles of the distributions; for any two $P, P' \in {\cal P}(\R)$ with quantiles $Q=F^{-1}, Q'=F^{' -1}$, respectively, the chosen Wasserstein metric would be
\begin{eqnarray*}
d_{M_2}(P,P')=\bigg( \int_{0}^{1} (Q(s)-Q'(s))^2 ds  \bigg)^{1/2}
=\bigg( \int_{0}^{1} (F^{-1}(u)-F^{' -1}(u))^2 du  \bigg)^{1/2}.
\end{eqnarray*}
The corresponding Fr\'echet mean is well defined and unique (if at least one of the distributions is absolutely continuous with respect to the Lebesgue measure see \cite{agueh2011barycenters}) so that
\begin{eqnarray*}
P_{B}=\arg\min_{P \in {\cal P}(\R)} \bigg(  \sum_{i=1}^{N} w_{i} d_{M_2}^2(P_i, P) \bigg),
\end{eqnarray*}
which leads to a representation for $P_{B}$ represented by a distribution function $F_{B}$ defined by the quantile averaging scheme
\begin{eqnarray}\label{APP-102}
F_{B}^{-1} =\sum_{i=1}^{N}w_{i} F_{i}^{-1} \,\,\, \Longleftrightarrow \,\,\, F_{B}=\bigg(  \sum_{i=1}^{N}w_{i} F_{i}^{-1} \bigg)^{-1}.
\end{eqnarray}
The representation \eqref{APP-102} is clearly a nonlinear representation for the mean, unlike the linear mean
\begin{eqnarray*}
\bar{F}=\sum_{i=1}^{N}w_{i} F_{i},
\end{eqnarray*}
which is commonly used in model averaging or consensus formation.

In case where the future risk is not represented in a satisfactory way by a single number, we can consider $M_2$ as the space of probability measures on $\R^d$, denoted by ${\cal P}(\R^d)$. This is not a vector space also, and can be metrized again using the Wasserstein metric (defined in Example \ref{WASSERSTEIN-METRIC-EX}). In this setup there are in general no closed form solutions for the Wasserstein metric. However, the Wasserstein barycenter, is well defined and unique if at least one of the measures $P_i$ is absolutely continuous with respect to the Lebesgue measure on $\R^{d}$, so that
\begin{eqnarray*}
P_{B}=\arg\min_{P \in {\cal P}(\R^d)} \bigg(  \sum_{i=1}^{N} w_{i} d_{M_2}^2(P_i, P) \bigg).
\end{eqnarray*}
In the special case of Location-Scatter families, where each probability measure is parametrized as $P_{i}=LS(\mu_i, S_{i})$, where $\mu_i \in \R^{d}$, $S_{i} \in M^{d\times d}_{+}$ the Wasserstein distance between any two measures $P_1,P_2$ is given by \eqref{APP-200}, and the corresponding Wasserstein barycenter is a probability measure from the same family 
\begin{eqnarray}\label{501}
Q_{B}=LS(\mu_B.S_{B}),
\end{eqnarray}
with covariance matrix satisfying the equation
\begin{eqnarray}\label{500}
S_{B}=\sum_{i=1}^{N} w_{i} (S_{B}^{1/2} S_{i} S_{B}^{1/2}).
\end{eqnarray}
Clearly this is far from the standard linear averaging on the vector space of in which distribution functions are naturally embedded.

The full opinion space in the social discount factor example is $M_{1}\times M_{2}$, endowed with the metric 
\begin{eqnarray*}
d((f,P),(f',P'))=d_{M_1}(f,f') + d_{M_2}(P,P').
\end{eqnarray*}
As mentioned above, the Fr\'echet mean is not always uniquely defined. In the examples used here, i.e. the Wasserstein case where at least one of the probability measures are absolutely continuous with respect to the Lebesgue measure, or for certain examples of manifolds, the uniqueness of the Fr\'echet mean has been established (see e.g. \cite{arnaudon2012medians}, \cite{arnaudon2014means}, \cite{agueh2011barycenters}). However, the proposed methodology is not directly affected if the chosen opinion space is such that the uniqueness of the Fr\'echet mean is not guaranteed. In such cases, a local minimizer for the Fr\'echet functional could serve very well as a local consensus point, and the proposed scheme could easily be applied for the determination of local consensus points. The situation can be considered analogous to the situation concerning Nash equilibria, which may not be unique, but nonetheless even local Nash equilibria can be a useful concept in decision making. 

\subsection{Algorithms for determining the Fr\'echet barycenter}

The algorithmic determination of the Fr\'echet barycenter depends crucially on the choice of the relevant metric space setup. We comment here for the choice of metric space, for the example of the consensus concerning the social discount rate, using the same notation as in the previous subsection. 

For the $M_1$ part we use the finite dimensional representation of the relevant set of curves ${\cal R}$, in order to calculate the Fr\'echet barycenter. In particular, given a set of yield curves $\{f_1, \cdots, f_N\} \subset {\cal R}$, we use their parameterization in terms of the parameters $\{\theta_1, \ldots, \theta_N\} \subset \Theta$, and express the (infinite dimensional) optimization problem 
\begin{eqnarray}\label{APP-301}
f_{B}=\arg\min_{f \in M_1} \bigg( \sum_{i=1}^{N} w_i d^2(f, f_i ) \bigg) ,
\end{eqnarray}
in terms of the finite dimensional optimization problem
\begin{eqnarray}\label{APP-302}
\theta_{B}=\arg\min_{\theta \in \Theta} \sum_{i=1}^{N}w_{i} \hat{d}^2(\theta, \theta_i),
\end{eqnarray}
which then characterizes $f_{B}$ in terms of
\begin{eqnarray*}
f_{B}= \Phi(\cdot, \theta_{B}).
\end{eqnarray*}
Problem \eqref{APP-302} is treated using standard finite dimensional techniques, e.g. first order optimization methods, or standard higher order methods, e.g. quasi-Newton methods. For the  numerical experiments presented here we have used a mixture of gradient based methods and quasi-Newton methods for the study of problem \eqref{APP-302} leading to the determination of the Fr\'echet mean. 

For the $M_2$ part, in the case where the loss distribution is one dimensional, the determination of the barycenter comes directly from the representation \eqref{APP-102}. The only demanding task from the point of view of computation in \eqref{APP-102} is the determination of the relevant quantiles. However, for the application in mind, a suitable class of distributions is the class of generalized extreme value distributions as in \eqref{400}, for which the quantiles are available in closed form analytically, and for which the quantile averaging
\eqref{APP-102} reduces to simple parameter averaging. In the case where the loss distribution is multi dimensional, we may restrict our attention to the class of Location Scatter family, and imply the representation of the Wassersstein barycenter in terms of the representation \eqref{501}, with $S_{B}$ given by \eqref{500}. Then, the only computationally demanding part is the determination of $S_{B}$, in terms of the solution of the matrix equation \eqref{500}. This can be easily performed numerically in terms of the fixed point scheme 
\begin{eqnarray}\label{503}
S_{B,k+1}=\sum_{i=1}^{N} w_{i} (S_{B, k}^{1/2} S_{i} S_{B, k}^{1/2}).
\end{eqnarray}
which quickly converges to the required value of the covariance matrix $S_{B}$, corresponding to the Wasserstein barycenter.

\bibliographystyle{chicago}
\bibliography{references}
\end{document}